\documentclass{ieeeaccess}
\usepackage{cite}
\usepackage{amsmath,amssymb,amsfonts}
\usepackage{algorithmic}
\usepackage{graphicx}
\usepackage{textcomp}
\usepackage{algorithm}
\usepackage{subcaption}

\def\BibTeX{{\rm B\kern-.05em{\sc i\kern-.025em b}\kern-.08em
    T\kern-.1667em\lower.7ex\hbox{E}\kern-.125emX}}
\begin{document}
\history{Date of publication xxxx 00, 0000, date of current version xxxx 00, 0000.}
\doi{10.1109/TQE.2020.DOI}

\title{Scheduling Concurrent Entanglement Requests in Quantum Networks}

\author{\uppercase{Gongyu Ni}\authorrefmark{1}, \uppercase{Lester Ho}\authorrefmark{1}}




\address[1]{Wireless Communications Laboratory, Tyndall National Institute, Dublin D02 W272, Ireland}


\markboth
{Author \headeretal: Preparation of Papers for IEEE Transactions on Quantum Engineering}
{Author \headeretal: Preparation of Papers for IEEE Transactions on Quantum Engineering}

\corresp{Corresponding author: Gongyu Ni (email: gongyu.ni@tyndall.ie)}

\begin{abstract}


This paper investigates resource allocation for entanglement distribution in multi-node, multi-channel quantum networks at the metropolitan scale. A multi-slot quantum network simulation framework is developed across physical and network layers, incorporating heterogeneous link characteristics, limited quantum memories, request queuing, retry mechanisms, and concurrent entanglement distribution. Based on this framework, a centralized scheduling architecture is proposed to coordinate quantum memories, communication channels, and routing paths for multiple simultaneous entanglement requests. Three classes of resource allocation strategies are evaluated: heuristic approaches, a mixed-integer linear programming (MILP) optimization method, and a Proximal Policy Optimization (PPO)-based reinforcement learning approach. The heuristic schemes reveal fundamental trade-offs between request delay and entanglement success rate: Dynamic Efficient minimizes delay, Success Enhancement improves success probability through adaptive multi-path allocation, and Static Efficient provides a balance between these objectives. The MILP approach achieves optimized resource allocation by jointly considering request handling and multi-path assignment, while the PPO-based method learns adaptive scheduling policies to improve overall performance. Simulation results demonstrate that these approaches provide different performance advantages in terms of request delay, entanglement success rate, capacity utilization, and request handling rate, highlighting the trade-offs between efficiency and reliability in metropolitan-scale quantum network scheduling.

\end{abstract}


\titlepgskip=-15pt

\maketitle

\section{Introduction}
\label{sec:introduction}

Quantum networks are an emerging communication technology designed to transmit quantum information between interconnected nodes. These nodes support a variety of applications, including quantum key distribution, distributed quantum computing, and quantum sensing. The fundamental building block of a quantum network is the generation of entanglement between nodes, enabling quantum correlations that facilitate communication.

Although entanglement has been demonstrated over a spatial separation of $1.3$~kilometers \cite{hensen2015loophole}, long-distance quantum networking is fundamentally limited by exponential photon loss in optical fibers \cite{pirandola2017fundamental}. Consequently, quantum repeater architectures employing multiplexed architecture have been proposed to enhance entanglement distribution rates \cite{lee2022quantum, muralidharan2016optimal, dai2021entanglement}. 

With the development of quantum repeaters, multi-node interconnected quantum networks have become an active area of research. One of the fundamental challenges in such networks is the routing problem, which aims to determine the optimal sequence of intermediate nodes for establishing end-to-end entanglement between two nodes.

Furthermore, the emergence of multi-channel quantum networks, enabled by wavelength-division multiplexing (WDM) techniques in quantum communication channels, has significantly increased the capacity of quantum networks \cite{qin2026monolithic,qiu2026integrated,wengerowsky2018entanglement}. Beyond multi-channel communication, multi-user quantum networking, where entanglement requests are generated simultaneously by multiple end users, has also been experimentally demonstrated in multi-channel quantum networks \cite{wen2022realizing}. Distributed quantum computing (DQC) further intensifies this requirement, as it relies on remote entanglement between distributed quantum processors to implement non-local quantum operations \cite{main2025distributed}. As DQC systems scale, multiple DQC jobs are expected to execute concurrently \cite{ferrari2024execution,ni2026advanced}, making efficient entanglement distribution increasingly important for supporting scalable distributed quantum computing.
Therefore, investigating routing and resource allocation under practical resource constraints, including limited quantum processing units, finite quantum memory, and constrained communication channels, has become a critical problem for the near-term quantum networks.


In this paper, we consider a multi-node, metropolitan-scale quantum network in which each physical link supports a fixed, limited number of simultaneous entanglement-generation attempts. These parallel attempts are assumed to be enabled by multi-channel techniques, such as wavelength-division multiplexing. The primary objective is to compare different scheduling strategy paradigms for concurrent entanglement requests. The main contributions of this paper are as follows:

\begin{itemize}

    \item We develop a quantum network simulation framework spanning the physical and network layers~\cite{pirker2019quantum,pompili2022experimental}. At the physical layer, we model limited quantum memories, heterogeneous photon loss rates, varying link distances, and dephasing and decoherence errors. At the network layer, a multi-slot simulation framework captures multiple request generation, queuing, retries, and execution over several network topologies.

    \item We propose a centralized scheduling framework that jointly supports concurrent entanglement distribution for multi-requests. The scheduler apply resource allocation strategies to coordinate the allocation of quantum memories, communication channels, and routing paths.

    \item We design and evaluate three classes of resource allocation strategies: (i) heuristic schemes, including Dynamic Efficient, which prioritizes minimizing request delay; Success Enhancement, which maximizes the entanglement success rate through adaptive multi-path allocation at the cost of higher delay; and Static Efficient, which provides a compromise by achieving lower delay than Success Enhancement while attaining a higher success rate than Dynamic Efficient; (ii) a mixed-integer linear programming (MILP) approach that computes an optimal resource allocation by jointly maximizing the request handle rate to reduce average delay and allocating multiple paths to individual requests to improve the entanglement success rate; and (iii) a Proximal Policy Optimization (PPO)-based reinforcement learning approach that learns adaptive resource allocation policies through reward-driven optimization.
    
\end{itemize}

\section{Related Work}
\label{sec: related_work}


In this section, we survey existing research on quantum network routing and entanglement request scheduling, with emphasis on heuristic and mixed-integer programming approaches for near-term quantum networks supporting multipath entanglement distribution.

Regarding the routing problem, under restrictions imposed by specific network topologies, Schoute et al.~\cite{schoute2016shortcuts} proposed a framework for studying routing in quantum networks. However, their analysis is limited to ring and spherical topologies. Pant et al.~\cite{pant2019routing} proposed routing solutions for entanglement distribution in grid networks. In addition, Chakraborty et al.~\cite{chakraborty2019distributed} studied the performance of routing algorithms over ring, grid, and recursively generated network topologies.

Beyond routing, the scheduling of newly arrived and pending requests in each time slot are also important. Cicconetti et al.~\cite{cicconetti2021} proposed a heuristic-based scheduling framework designed to minimize delay while optimizing entanglement generation rate and fidelity. Le and Nguyen~\cite{le2022dqra} focused on routing requests in resource-constrained quantum networks by introducing the Deep Quantum Routing Agent (DQRA), a deep reinforcement learning model that enhances entanglement routing paths to maximize request handling capability. Ni et al.~\cite{ni2025entanglement} applied a Deep Q-Network (DQN)-based scheduling scheme to balance the conflicting objectives of minimizing delay and maximizing fairness among entanglement requests.

For heuristic algorithms in multipath entanglement distribution, Shi et al.~\cite{shi2020concurrent} proposed a routing algorithm that improves the success probability of long-distance entanglement generation by dynamically selecting candidate paths based on local link-state information. Their approach assumes that each node has knowledge of the link states of its neighbouring nodes. However, acquiring and updating neighbouring link-state information introduces additional communication overhead and latency, which may be undesirable in quantum networks due to the limited coherence time of quantum memories. 
Li et al.~\cite{li2021effective} proposed a progressive-filling algorithm based on propagatory updates and evaluated it on lattice networks with homogeneous links. In contrast, we develop and evaluate allocation methods for multiple random topologies with heterogeneous link characteristics.
More recently, Halder et al.~\cite{halder2024concurrent} proposed a concurrent multipath entanglement distribution framework for serving multiple simultaneous entanglement requests through joint path selection and resource allocation. Their heuristic approach improves the success probability and resource utilization by exploiting multiple candidate paths under limited quantum resources. However, their evaluation considers a single scheduling instance with a batch of concurrent requests, without investigating continuous multi-slot scheduling under dynamic request arrivals and resource evolution. In contrast, our work evaluates scheduling algorithms over multiple time slots, providing a more realistic assessment of long-term network performance.

For mixed-integer and linear programming (MILP) approaches, Vardoyan et al.~\cite{vardoyan2024bipartite} formulated a mixed-integer quadratically constrained program to optimize bipartite entanglement distribution, where multiple entangled pairs can be established between a single source--destination pair through multiple available paths. To address concurrent entanglement distribution among multiple source--destination pairs, Chakraborty et al.~\cite{chakraborty2020entanglement} proposed a linear programming (LP) formulation that maximizes the aggregate entanglement distribution rate while satisfying end-to-end fidelity constraints. However, their approach focuses primarily on maximizing entanglement generation rate of the network rather than satisfying individual entanglement demands from users. From a user-request perspective, Zeng et al.~\cite{zeng2022multi} formulated multi-user entanglement throughput optimization as a two-stage integer programming problem, aiming to maximize the expected throughput among user pairs. Nevertheless, their model mainly considers fiber transmission and successful entanglement swapping probability as the determinants of end-to-end entanglement success, without incorporating other quantum factors. Therefore, our MILP formulation jointly addresses entanglement demand fulfilment and the maximization of final entanglement request success probability.

\section{System Model} 
\label{sec: system_model}

The integrated system model incorporates a quantum model that captures depolarizing and dephasing error, as well as photon loss during propagation. It also includes a network model that characterizes request arrivals, pending, and retry mechanisms within a multi-slot simulation framework.

\subsection{Quantum model}
\label{subsec: quantum_model}
In the quantum model, Bell pairs $|\Phi\rangle$ (Eq.~(\ref{eqn:bell_pair})) are used to build entanglement between two nodes. 

\begin{equation}
\label{eqn:bell_pair}
|\Phi\rangle=\frac{1}{\sqrt{2}}\left(|0\rangle|0\rangle+|1\rangle|1\rangle\right)
\end{equation}

In a directly connected two-node system, Bell pairs are distributed between the two nodes.
When intermediate nodes (repeaters) exist between the end nodes, quantum operations such as entanglement swapping and correction are employed to establish end-to-end entanglement \cite{van2013}. This process involves three main steps: (1) the intermediate node establishes entanglement with adjacent neighbours, (2) it performs a Bell-state measurement to swap entanglement, and (3) the measurement outcomes are transmitted via classical channels to the end nodes, which apply appropriate $X$ and $Z$ corrections to their local qubits. To simplify the multi-node chain entanglement, repeaters that are within the path of two end nodes, are treated equivalently to end nodes, forming a network composed of nodes and quantum channels. 

To evaluate the success of distributing entanglement, fidelity $F$ is used to measure the difference between the received density matrix $\rho$ and the expected state $|\psi\rangle$ \cite{van2014}.

\begin{equation}
F(|\psi\rangle, \rho) =\langle\psi|\rho| \psi\rangle
\end{equation}

To model photon loss during propagation, the photon loss probability $p_{\text{loss}}$ over a channel is computed using Eq.~(\ref{eq:photon_loss}), where $d$ denotes the physical length of the link between two connected nodes. The fiber attenuation $r_{\text{loss}}$, representing classical optical loss, is randomly sampled from a predefined range ($0.15$ to $0.35$ dB/km), similar to the values reported for standard single-mode fibers in \cite{petrovich2025broadband}. This sampling is used during quantum network topology generation to model heterogeneous link characteristics.

\begin{equation}
\label{eq:photon_loss}
p_{\text{loss}} = 1 - 10^{-\frac{d \cdot r_{\text{loss}}}{10}}
\end{equation}

The quantum processor operates on qubits stored in the quantum memory to perform entanglement swapping, which is realized through a Bell-state measurement (BSM) on two qubits, followed by conditional Pauli corrections using $X$- and $Z$-gates according to the two classical correction bits. We assume single-qubit $X$-gates and $Z$-gates require $10\,\mathrm{ns}$ operation time. Since a Bell-state measurement consists of an entangling gate, single-qubit rotations, and qubit readout, we model its time duration as $500\,\mathrm{ns}$, which is consistent with experimentally reported superconducting qubit gate and readout latencies~\cite{bengtsson2024model}.

The probability of dephasing errors $p_{\text{dephase}}$ occurring in quantum memories is modelled as
\begin{equation}
\label{eq:dephase_error}
p_{\text{dephase}} = 1 - e^{-t_{\text{dephase}} \cdot r_{\text{dephase}}},
\end{equation}
where $t_{\text{dephase}}$ denotes the storage duration of a qubit in quantum memory, and $r_{\text{dephase}}$ represents the corresponding dephasing rate.

Similarly, depolarizing errors $p_{\text{depol}}$ during quantum operations are modelled as
\begin{equation}
\label{eq:depol_error}
p_{\text{depol}} = 1 - e^{-t_{\text{depol}} \cdot r_{\text{depol}}},
\end{equation}
where $t_{\text{depol}}$ denotes the duration of the quantum operation, and $r_{\text{depol}}$ represents the associated depolarization rate.

The depolarizing rate is set to $1$~kHz and the dephasing rate to $0.1$~MHz, following the minimal experimental parameters in \cite{cicconetti2021}. 

For entanglement generation, the quantum source is assumed to probabilistically generate Bell pairs with a source fidelity of $0.9$, i.e., the target Bell state is produced with probability $0.9$, while an erroneous state is generated with probability $0.1$. The entanglement generation rate is assumed to be fixed and identical for all links.

To model multi-channel entanglement establishment, we assume that all channels on each link are used simultaneously for individual entanglement processes. In our simulations, each link supports $5$ concurrent entanglement attempts in their available channels.

To model end-to-end entanglement establishment, path characteristics, including transmission distances among the number of intermediate nodes, together with quantum processes such as entanglement generation, quantum processor operations, and photon transmission loss, are simulated using NetSquid \cite{coopmans2021netsquid}. The simulator provides execution time, fidelity, and success status of each path. 

A fidelity threshold of $0.83$, corresponding to the theoretical threshold for Quantum Key Distribution (QKD) based on the Clauser--Horne--Shimony--Holt (CHSH) Bell inequality ($2(\sqrt{2}-1)\approx0.8284$), is used to determine whether entanglement establishment is successful \cite{sajeed2020bright}.
This value is selected as a representative minimum fidelity operating point for the simulations and is not intended to represent a universal threshold for QKD or other quantum-network applications, since the required fidelity depends on the specific protocol and application.
An entanglement request is considered successful only when its final fidelity is at least 0.83 and the entanglement-generation process is completed within the allocated time slot.

The quantum simulation parameters assumed in this work are summarized in Table~\ref{tab:quatum_parameters}.

\begin{table}
\caption{\label{tab:quatum_parameters}Parameters in quantum simulation.}
\setlength{\tabcolsep}{3pt}
\renewcommand{\arraystretch}{1.2}
\begin{tabular}{|p{130pt}|p{90pt}|}
\hline
Description & 
Value \\
\hline
Target state generation probability & 
0.9 \\
Depolarizing rate in quantum memory & 
1 KHz \\
Quantum operation dephasing rate & 
0.1 MHz \\
Single gate instruction time & 
$10\,\mathrm{ns}$ \\
Bell-state measurement time & 
$500\,\mathrm{ns}$ \\
fiber attenuation selection set [dB/km]& 
{0.15, 0.2, 0.25, 0.3, 0.35} \\
Entanglement attempt per request & 
5 \\
Fidelity threshold & 
0.83 \\
\hline
\end{tabular}
\label{tab1}
\end{table}

\subsection{Network Model} \label{sec:networkmodel}

\begin{figure*}
  \centering
  \includegraphics[width=0.85\textwidth]{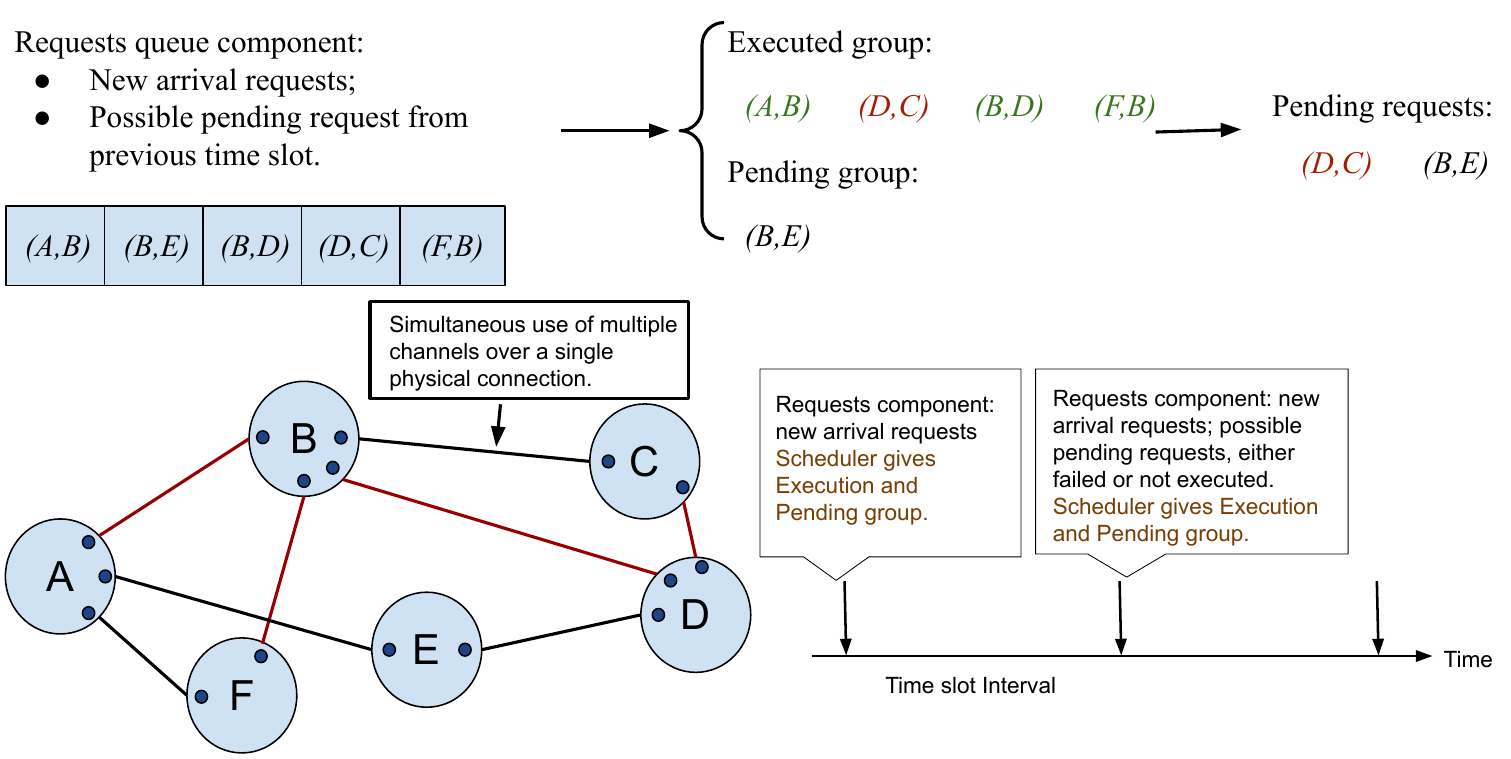}
  \caption{Centralized scheduling of entanglement requests in an example quantum network topology over multiple time slots. The requests $(A,B)$, $(D,C)$, $(B,D)$, and $(F,B)$, with their selected paths highlighted in red, are scheduled for parallel execution. The request $(D,C)$, however, fails to meet the required entanglement fidelity by the end of the time slot and is moved, along with $(B,E)$, to the pending group for the next slot. At each time slot, network resources are refreshed, making all feasible links available. The request set comprises both newly arrived and pending requests, after which a scheduler determines which requests are executed and which remain pending.
 }
  \label{fig:capacity_graph}
  \hrulefill
\end{figure*}

In the network model, as illustrated in Figure~\ref{fig:capacity_graph}, connectivity between two nodes is determined by the presence of a physical link (i.e. an existing optical fiber) and the requirement that each node is equipped with quantum memories for quantum state storage. This type of node has also been discussed in~\cite{lee2022quantum} for connecting neighbouring nodes. Moreover, the model considers multiple entanglement requests within each time slot, encompassing both newly arrived requests and those pending from the previous time slot.

When generating arriving requests, the source and destination nodes are chosen randomly. The number of arrivals $N_a$ follows a truncated Poisson distribution over the interval $[a,b]$, as given in Eq.~(\ref{eqn:possion}), where $a$ and $b$ denote the minimum and maximum number of requests per time slot, respectively.
\begin{equation}
N_a \sim \mathrm{Poisson}(\lambda)\big|_{[a,b]}
\label{eqn:possion}
\end{equation}

It is assumed that only one group of requests is executed in parallel per time slot, referred to as the executed group. Requests not selected for execution are placed in a pending group due to the unavailability of paths for establishing entanglement.

Upon executing the requests in the executed group, each request is classified as either successful or failed. Successful requests are defined as those that meet or exceed the fidelity threshold and are completed within a timeslot. Their delay, success status, and resulting fidelity are recorded. Failed requests are moved to the pending group and retried in subsequent time slots if they have not reached the maximum number of retries; otherwise, they are permanently aborted.

In summary, all requests are placed in a queue to be processed by the scheduler. As entanglement requests are executed immediately after the scheduler determines the resource allocation at the beginning of each time slot, the scheduling time does not affect the quantum process. Furthermore, since the entanglement generation rate is assumed to be fixed and identical for all links, the entanglement generation process is identical. At each time slot, the scheduler selects and executes requests, manages the pending queue, and discards requests that exceed the maximum allowed number of retries.

\section{Problem formulation}
\label{sec: prob_formulation}

In this section, we define the path cost metric and the path disjointness constraints for parallel scheduling. The path cost is formulated by incorporating photon loss, decoherence errors, and entanglement swapping cost along the path. Path disjointness is enforced because physical link resources, node locations, and quantum memory capacities remain fixed within each time slot and are replenished only at the beginning of the next time slot. Consequently, only requests with non-overlapping resource allocations can be scheduled for parallel execution.

\subsection{Path cost}

Based on the quantum entanglement distribution process, the path cost is designed to capture the impact of photon loss, quantum memory decoherence, and entanglement swapping success probability. In the path cost calculation, the entanglement swapping success probability is assumed to be $0.8$. Based on Eqs. (\ref{eq:photon_loss})-(\ref{eq:depol_error}), the cost components are defined as follows.

The transmission cost of a quantum link $(u,v)$ is defined as

\begin{equation}
\label{eq:loss_cost}
c_{uv} = -\log(1-p_{\mathrm{loss}}),
\end{equation}

where $p_{\mathrm{loss}}$ denotes the photon loss probability determined by the transmission distance and fiber attenuation.

Quantum memories introduce decoherence errors during qubit storage, including depolarization and dephasing. The corresponding memory cost is defined as

\begin{equation}
\label{eq:memory_cost}
c_{\mathrm{mem}} = -\log\left((1-p_{\mathrm{depol}})
(1-p_{\mathrm{dephase}})\right).
\end{equation}

The cost associated with entanglement swapping is defined as

\begin{equation}
\label{eq:swap_cost}
c_{\mathrm{swap}} = -\log(p_{\mathrm{swap}}),
\end{equation}

where $p_{\mathrm{swap}}$ is the success probability of an entanglement swapping operation.

Therefore, for a path $p=(v_0,v_1,\ldots,v_m)$, the total path cost is given by

\begin{equation}
\label{eq:path_cost}
C_p = \sum_{j=0}^{m-1} c_{v_jv_{j+1}} + S_p(c_{\mathrm{mem}}+c_{\mathrm{swap}}),
\end{equation}

where $S_p=m-1$ represents the number of entanglement swapping operations required along the path. A lower path cost indicates that the path provides a higher feasibility for establishing entanglement between the end nodes.

\subsection{Scheduling Parallelized Disjoint Paths}

As the entanglement generation rate is assumed to be identical for all physical links, the available quantum resources are replenished at a fixed rate. Consequently, the quantum network topology is represented by a static graph $G$ throughout each time slot. During every time slot, a central scheduler allocates these resources to a set of requests $\mathcal{R}$, consisting of newly arrived requests together with pending requests carried over from previous time slots.

Entanglement requests establish end-to-end entanglement along their assigned paths. Once a path is allocated by the scheduler, all links and their associated quantum resources are exclusively reserved for that request during the current time slot and cannot be shared with other requests. Consequently, all selected paths must be mutually edge-disjoint.

Given a network topology $G$, the set of feasible candidate paths for each request $r \in \mathcal{R}$ between the source--destination pair $(s_r, d_r)$ is defined as

\begin{equation}
\mathcal{P}_r=\{p_{r,1},p_{r,2},\ldots,p_{r,k}\},
\end{equation}

where each candidate path $p_{r,k}$ connects $s_r$ and $d_r$ through available physical links.

Allocating multiple edge-disjoint paths to a single request can improve the probability of successful entanglement establishment. However, each additional path consumes network resources that could otherwise serve other requests. Therefore, the scheduler should employ a resource allocation strategy that optimizes the trade-off between maximizing the number of served requests and allocating additional paths to improve the end-to-end entanglement generation success probability.

Let $\mathcal{P}'$ denote the set of all allocated paths in each time slot. To ensure exclusive resource usage, any two allocated paths must satisfy

\begin{equation}
E(p_f)\cap E(p_g)=\emptyset,
\qquad
\forall\, p_f,p_g\in\mathcal{P}',\; f\neq g,
\label{eq:edge_disjoint}
\end{equation}

where $E(p)$ denotes the set of physical links traversed by path $p$.

Furthermore, the allocated paths of every accepted request must satisfy

\begin{equation}
\mathcal{P}'_r\subseteq\mathcal{P}_r,
\qquad
\forall\, r\in\mathcal{R}',
\end{equation}

where $\mathcal{R}'\subseteq\mathcal{R}$ denotes the set of accepted requests and $\mathcal{P}'_r$ is the set of paths assigned to request $r$. Therefore, the resource allocation method is to determine the subset of executed requests $\mathcal{R}'$ together with a feasible path assignment that satisfies the edge-disjointness constraint while allowing multiple disjoint paths to be allocated to a request whenever sufficient resources are available.

The general resource allocation pseudocode in Protocol~\ref{alg:general_allocation} outlines steps for resource updating and allocation, including methods for selecting parallel disjoint paths.

\begin{algorithm}
\caption{General Allocation Protocol}
\label{alg:general_allocation}
\begin{algorithmic}[1]

\STATE \textbf{Step 0: Initialization}
\STATE Renew all available links in the quantum network.
\vspace{0.4em}
\STATE \textbf{Step 1: Request Collection}
\STATE Obtain the set of requests $\mathcal{R}$ within the current time slot.
\vspace{0.1em}
\STATE \textbf{Step 2: Allocation Scheduling Algorithms}
\STATE \textbf{Input:} Graph $G$, Set of requests $\mathcal{R}$
\STATE Selects requests based on relevant attributes (e.g., source and destination nodes, path length, path cost, required quantum memory, and number of requests).
\STATE \textbf{Output:} Selected requests with assigned paths; unselected requests are added to the pending group.
\vspace{0.4em}
\STATE \textbf{Step 3: Entanglement and Evaluation}
\STATE Establish entanglement between end nodes along the assigned paths and evaluate the outcomes.
\vspace{0.4em}
\STATE \textbf{Step 4: Retry Mechanism}
\STATE Failed requests are moved to the pending group for the next time slot according to the retry mechanism.

\end{algorithmic}
\end{algorithm}

\section{Evaluation Metrics}
\label{sec: eva_metrics}

To evaluate the performance of different allocation methods, we adopt the following metrics: delay time in successful requests, total number of successfully served entanglement requests over the entire simulation, capacity utilization (measuring link resource usage), and handling rate (reflecting request processing efficiency).

\subsection{Delay time}

The delay time for successful request, $t_d$, in Eq. (\ref{eqn:request_delay_time}), is determined by subtracting the time $t_{f}$ at which the request was fulfilled from the time $t_{g}$ at which it was generated.

\begin{equation}  
\label{eqn:request_delay_time}  
    t_{d} = t_{f} - t_{g}  
\end{equation}      

\subsection{Number of successful requests}

The number of successes $n_s$ is defined as the number of successfully established entanglement requests throughout the simulation.

\subsection{Capacity Utilization}

The capacity utilization rate $u$ in Eq. (\ref{eqn:utilization}), is the ratio of the total number of links used per slot $l_u$ to the total number of links $L$ available in the network topology. A higher value indicates that more links are utilized per slot.

\begin{equation}
\label{eqn:utilization}
u = \frac{l_{u}}{L}
\end{equation}

\subsection{Handling rate}

The handling rate $r_h$, defined in Eq.~(\ref{eqn:handle_rate}), is the ratio of the number of executed requests $n_e$ to the total number of requests $n_r$ per time slot. A higher handling rate indicates that the scheduler is more capable of executing entanglement requests in parallel.

\begin{equation}
\label{eqn:handle_rate}
r_h = \frac{n_e}{n_r}
\end{equation}


\section{Resource allocation methods}
\label{sec: scheduling_methods}

In this section, we consider three classes of resource allocation strategies:
(i) heuristic methods, including Dynamic Efficient, Static Efficient, and Success Enhancement schemes;
(ii) a mixed-integer linear programming (MILP) formulation that derives optimal resource allocations by solving an explicit optimization problem; and
(iii) a reinforcement learning (RL) approach based on Proximal Policy Optimization (PPO) to learn adaptive allocation policies through reward-based optimization.

\subsection{Dynamic Efficient Allocation}

The goal of the Dynamic Efficient allocation method (Algorithm \ref{alg:dynamic_efficient_allocation}) is to accommodate as many entanglement requests as possible within each time slot, aiming to handle most of the requests. 

The steps involved are as follows.
First, the lowest-cost path that requires the lowest quantum memories is assigned for a given request, in comparison with all other requests. This strategy minimizes the impact on subsequent requests, as a low-cost path is typically shorter and has a higher probability of success. Next, the links of the selected path are removed from the graph, resulting in a sub-graph. Based on this sub-graph, the steps above are repeated for the next requests. This process continues until all requests are accommodated or no additional requests can be added due to resource limitations. 

\begin{algorithm}
\caption{Dynamic Efficient Allocation Algorithm}
\label{alg:dynamic_efficient_allocation}
\begin{algorithmic}[1]
\STATE \textbf{Input:} Graph $G$, Set of requests $\mathcal{R}$
\STATE \textbf{Output:} Dictionary of selected requests

\STATE // Evaluate requests
\FOR{each request in the list}
    \STATE Find candidate paths between request's source and destination
    \IF{valid paths exist}
        \STATE Select the best path based on cost and required memories
        \STATE Append evaluation result to the list
    \ENDIF
\ENDFOR
\STATE Sort evaluations by cost and required memories

\STATE // Process requests
\STATE Initialize graph and step counter
\WHILE{requests remain}
    \STATE Evaluate requests with current graph
    \IF{no valid evaluations left} \STATE break \ENDIF
    \STATE Select the best evaluation
    \STATE Store selected evaluation for current step
    \STATE Remove the selected path from the graph
    \STATE Remove the selected request from the list
\ENDWHILE
\STATE \textbf{Return} groups of selected and unselected requests
\end{algorithmic}
\end{algorithm}

\subsection{Static Efficient Allocation}

For the Static Efficient Allocation method, the network graph and the corresponding paths for the requests are not updated. This method achieves minimal computational complexity by first computing the lowest-cost path for each request and sorting all requests in ascending order of path cost. Requests are then combined sequentially, starting with the request with the minimum-cost path. When the next candidate path cannot be accommodated together with the already selected paths due to resource constraints, the process terminates. The resulting set of compatible requests is scheduled to be executed in parallel within the current time slot.

\begin{figure*}
\centering
\includegraphics[width=0.9\textwidth]{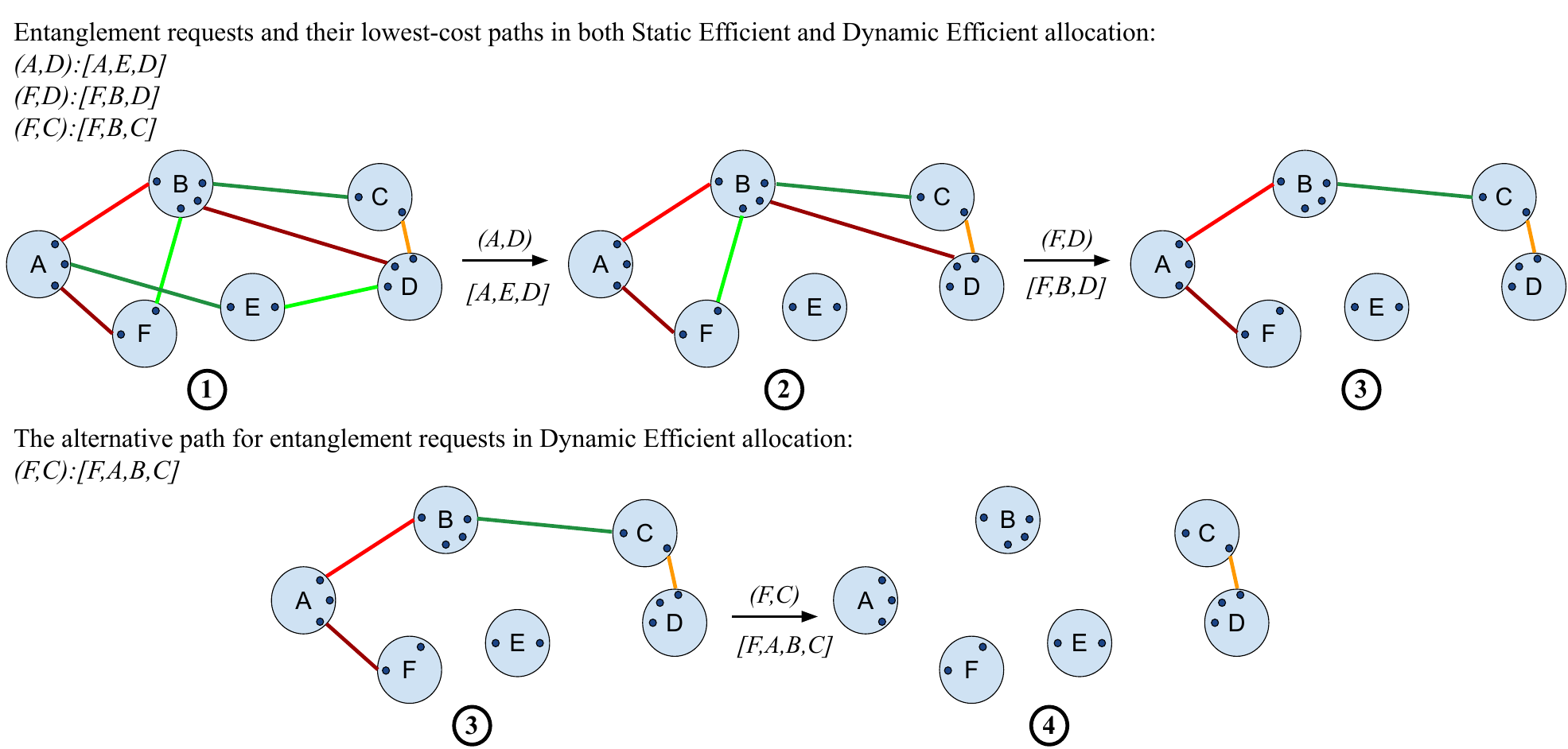}
\caption{Comparison of path allocation for sequential entanglement requests $(A,D)$, $(F,D)$, and $(F,C)$.
In the static efficient allocation, paths are predetermined, preventing request $(F,C)$ from being served in parallel at step~3 because its predefined lowest-cost path, $[F,E,C]$, is unavailable.
In contrast, the dynamic efficient allocation iteratively selects paths on the updated subgraph, where the alternative path $[F,A,B,C]$ remains available, enabling $(F,C)$ to be served in parallel with the other requests. Warmer colours indicate higher photon loss rates.}
\label{fig:dynamic_static}
\hrulefill
\end{figure*}

Compared with the Dynamic Efficient allocation shown in Figure~\ref{fig:dynamic_static}, the Static Efficient allocation does not update paths based on the latest sub-graph. Although the Dynamic Efficient allocation can process more entanglement requests in parallel, it may assign higher-cost paths to requests near the end of the queue. In contrast, the Static Efficient allocation guarantees that each request is served on its optimal path, at the cost of potential delays when requests are deferred to subsequent time slots.

\subsection{Success Enhancement Allocation}

In the Success Enhancement allocation method (Algorithm~\ref{alg:success_enhancement}), requests are classified into three categories: good, medium-worst and worst, based on their lowest path cost. Requests with a minimum path cost below the threshold $\tau_g$ are classified as good, those with a cost exceeding the threshold $\tau_w$ are classified as worst, and the remaining requests are categorized as medium-worst.

During each time slot, resource allocation prioritizes the medium-worst requests by executing them first. For these requests, the $k$th lowest-cost paths are considered for allocation only if their costs are comparable to the lowest-cost path, i.e., within a similarity threshold $\tau_s$. This multipath strategy increases the probability of successful entanglement establishment: if one path fails, alternative paths may succeed, and once any path successfully establishes entanglement, the corresponding end nodes are entangled.

In contrast, good requests are generally easier to fulfil, while worst requests typically require more time and exhibit a higher probability of failure. Therefore, only a single path is allocated to these two categories. This approach conserves resources for good requests while avoiding excessive resource consumption and time wastage for worst requests.

In this paper, the path cost thresholds are set to $\tau_g = 0.6$ and $\tau_w = 1.2$ to classify requests into good, medium, and worst categories. Under the simulation settings, the link cost $c_{uv}$ ranges from $0.17$ to $1.21$, while the memory and swap costs are approximately $c_{\mathrm{mem}} = 0.05$ and $c_{\mathrm{swap}} = 0.22$, respectively. Consequently, the cost associated with a single repeater operation, including both memory and entanglement swapping, is approximately $0.27$. Based on these values, path costs exceeding $1.2$ are classified as challenging for successful entanglement establishment, as such costs can arise from paths spanning more than three nodes and traversing links of moderate quality. In contrast, path costs below $0.6$ are considered easier to build entanglement, as they typically correspond to paths requiring fewer than two repeaters and traversing relatively high-quality links. In addition, the path cost similarity threshold is set to $\tau_s = 0.2$ and the maximum number of candidate paths is restricted to $k=2$.

\begin{algorithm}
\caption{Success Enhancement Algorithm}
\label{alg:success_enhancement}
\begin{algorithmic}[1]
\STATE \textbf{Input:} Graph $G$, Set of requests $\mathcal{R}$
\STATE \textbf{Output:} Dictionary of selected requests

\STATE // Find requests for allocating multiple paths
\FOR{each request in the list}
    \STATE Find multiple paths between source and destination
    \IF{multiple paths exist}
        \STATE Select valid paths based on cost and parallization
        \STATE Store valid paths
    \ENDIF
\ENDFOR

\STATE // Evaluate requests
\FOR{each request in valid paths}
    \STATE Calculate the total cost of paths
    \IF{cost is within acceptable range}
        \STATE Store evaluated request
    \ENDIF
\ENDFOR
\STATE Sort requests by cost

\STATE // Process requests
\STATE Initialize new graph copy and step counter
\STATE Initialize selected requests list
\WHILE{requests remain}
    \STATE Select and evaluate paths from requests
    \IF{valid path exists in the graph}
        \STATE Allocate path and update graph
        \STATE Remove request from remaining list
        \STATE Store selected request in current step
    \ENDIF
\ENDWHILE
\STATE \textbf{Return} groups of selected and unselected requests
\end{algorithmic}
\end{algorithm}

\subsection{MILP Allocation}

To solve the resource allocation problem, a Mixed Integer Linear Programming (MILP) formulation is developed. The formulation determines the optimal set of requests and corresponding paths while satisfying the edge-disjoint resource constraints.

\subsubsection{Decision Variables}

For each request $r\in\mathcal{R}$ and candidate path $p_{r,k}\in\mathcal{P}_r$, define the binary variable:

\begin{equation}
x_{r,k}=
\begin{cases}
1, & \text{if path }p_{r,k}\text{ is allocated to request }r,\\
0, & \text{otherwise}.
\end{cases}
\end{equation}

A second binary variable is introduced to indicate whether request $r$ is selected for execution:

\begin{equation}
y_r=
\begin{cases}
1, & \text{if request }r \text{ is in execution group},\\
0, & \text{otherwise}.
\end{cases}
\end{equation}

\subsubsection{Objective Function}

The objective is to maximize the number of requests executed in parallel while assigning additional paths to selected requests to improve their end-to-end entanglement success probability. To balance these objectives, the following weighted objective function is defined:

\begin{equation}
\max
\left(M\sum_{r\in\mathcal{R}}y_r + \sum_{r\in\mathcal{R}}\sum_{k\in\mathcal{P}_r}x_{r,k}\right),
\label{eq:milp_objective}
\end{equation}

where $M$ is a constant that balances the trade-off between the first and second terms. The first term is prioritized to maximize throughput, while the second term maximizes the number of additional paths allocated to accepted requests. In this paper, $M$ is set to $10$ to highlight the impact of the request handling rate. To ensure a fair comparison with the Success Enhancement allocation method, $k$ is set to $2$, allowing two candidate lowest-cost paths to be considered for each request.

\subsubsection{Request Service Constraint}

A request is considered served if at least one candidate path is selected:

\begin{equation}
\sum_{k\in\mathcal{P}_r}x_{r,k}
\geq y_r,
\quad
\forall r\in\mathcal{R}.
\label{eq:service_constraint}
\end{equation}

Moreover, a candidate path can be selected only when the corresponding request is selected for execution:

\begin{equation}
x_{r,k}\leq y_r,
\quad
\forall r\in\mathcal{R},
\forall k\in\mathcal{P}_r .
\label{eq:path_selection_constraint}
\end{equation}

These constraints allow each request in the execution group to be assigned one or multiple paths.

\subsubsection{Link-disjoint Resource Constraint}

Let the binary parameter $\delta_{r,k}^{e}$ indicate whether edge $e$ is included in candidate path $p_{r,k}$, defined as

\begin{equation}
\delta_{r,k}^{e} =
\begin{cases}
1, & \text{if } e \in p_{r,k},\\
0, & \text{otherwise}.
\end{cases}
\end{equation}

Since an allocated path removes all its occupied links from the available network, each physical link can only be assigned to one selected path:

\begin{equation}
\sum_{r\in\mathcal{R}}
\sum_{k\in\mathcal{P}_r}
\delta_{r,k}^{e}x_{r,k}
\leq 1,
\quad
\forall e\in E .
\label{eq:edge_constraint}
\end{equation}

This constraint guarantees that all selected paths are mutually edge-disjoint.

\subsubsection{Binary Constraints}

The decision variables satisfy:

\begin{equation}
x_{r,k}\in\{0,1\},
\quad
\forall r\in\mathcal{R},
k\in\mathcal{P}_r ,
\end{equation}

\begin{equation}
y_r\in\{0,1\},
\quad
\forall r\in\mathcal{R}.
\end{equation}

The resulting MILP jointly optimizes request admission and multipath resource allocation. After solving the MILP, the selected paths are removed from the current graph to update the network state for the next time slot.

\subsection{Proximal Policy Optimization allocation}

\begin{figure}
\centering
\includegraphics[width=0.9\linewidth]{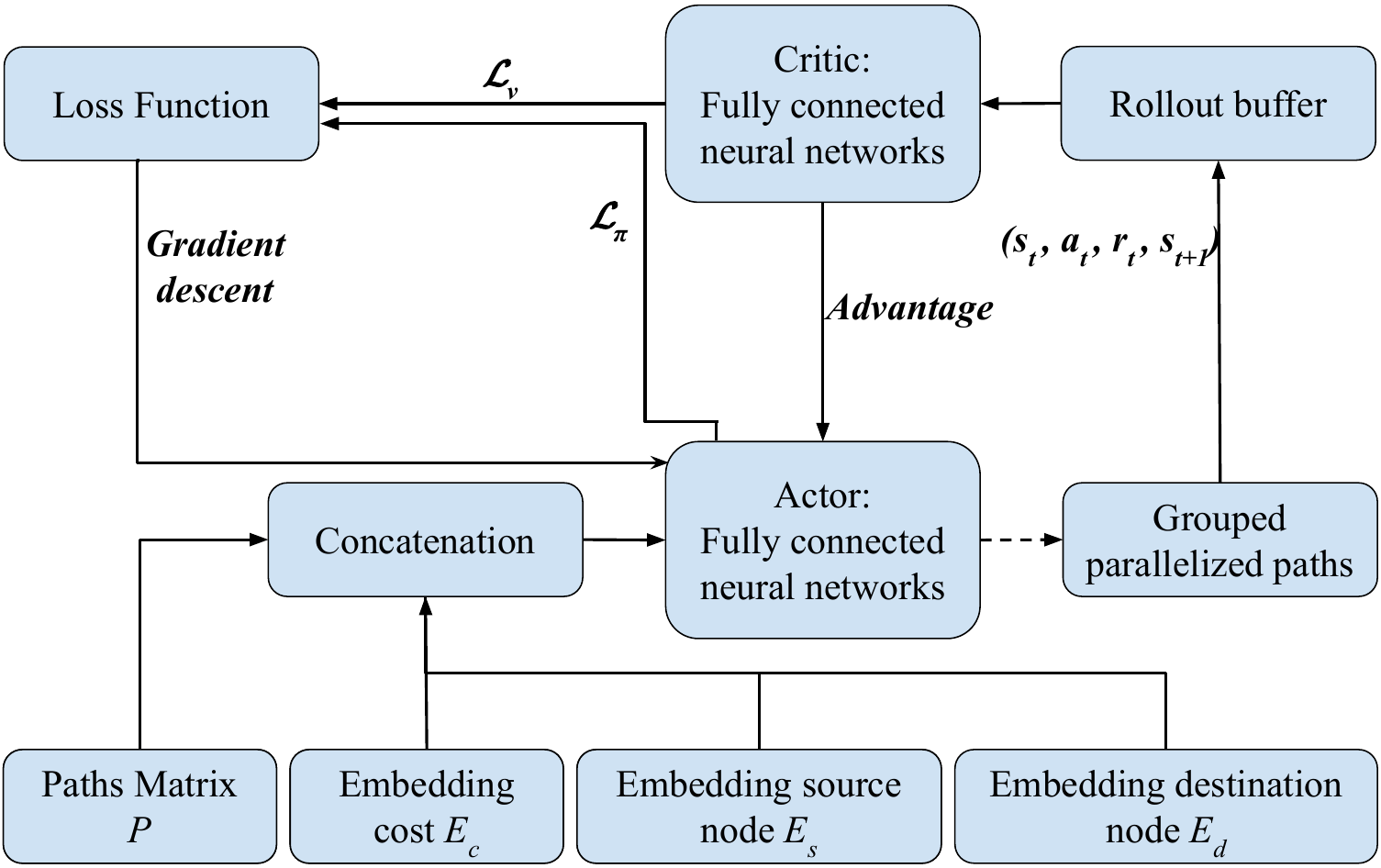}
\caption{The PPO model structure for the multipath quantum network.}
\label{fig:ppo_structure}
\hrulefill
\end{figure}

Proximal Policy Optimization (PPO) \cite{schulman2017proximal} is employed in this work as a representative learning-based scheduling approach. The objective is to evaluate the trade-offs of learning-based allocation relative to heuristic and MILP-based methods, rather than to establish PPO as the best reinforcement-learning algorithm for quantum-network scheduling. PPO is selected because it supports sequential policy-based decision-making, training stability and robustness to hyperparameter selection. In this work, we integrate PPO into the proposed quantum resource allocation framework by defining the state representation, action space, and reward function, as illustrated in Fig.~\ref{fig:ppo_structure}.

The specific definitions of the input state, action, reward, and loss function are given as follows:

\subsubsection{Input state}

The input state comprises the path matrix $P$, the cost embedding matrix $E_c$, and the source and destination embedding matrices $E_s$ and $E_d$. These components are formally defined as follows.

\paragraph{Paths Matrix}  
Multiple paths are selected for each request, where each path is represented as a binary vector indicating the presence of nodes. The paths matrix $P$, comprising a total of $n$ paths, is defined as:
\begin{equation}
P = 
\begin{bmatrix}
p_1 \\
p_2 \\
\vdots \\
p_{n}
\end{bmatrix}, \quad 
p_i = [v_1^{(i)}, v_2^{(i)}, \dots, v_{N}^{(i)}],
\end{equation}
where $N$ is the number of nodes, and each node $v_j^{(i)}$ is defined as:
\begin{equation}
v_j^{(i)} =
\begin{cases}
1, & \text{if path } p_i \text{ contains node } j, \\
0, & \text{otherwise}.
\end{cases}
\end{equation}

\paragraph{Cost Embedding.}  
Each path $p_i$ is associated with a calculated path cost $C_i$. Instead of using raw costs, a learnable embedding function $f_c(\cdot)$ is applied:
\begin{equation}
E_c = 
\begin{bmatrix}
f_c(C_1) \\
f_c(C_2) \\
\vdots \\
f_c(C_n)
\end{bmatrix}.
\end{equation}

\paragraph{Source and Destination Embeddings.}  
Each path $p_k$ is associated with a source-destination pair $(s_k, d_k)$, where $s_k, d_k \in \{1, \dots, N\}$ denote the indices of the source and destination nodes of the $k$-th path, respectively.  

The node indices are then mapped via learnable embedding functions, yielding the corresponding embedded representations:
\begin{equation}
E_s = 
\begin{bmatrix}
f_s(s_1) \\
f_s(s_2) \\
\vdots \\
f_s(s_n)
\end{bmatrix}, \quad
E_d = 
\begin{bmatrix}
f_d(d_1) \\
f_d(d_2) \\
\vdots \\
f_d(d_n)
\end{bmatrix},
\end{equation}
where each row corresponds to the embedding of the source or destination node associated with a path.

\paragraph{State Representation.}  
The final input state $s$ representation is constructed by concatenating all components for each path:
\begin{equation}
s = \{ P \;\; E_c \;\; E_s \;\; E_d \}.
\end{equation}


\paragraph{State zero-padding}  

As introduced in the network model, the number of requests varies across time slots due to dynamic arrivals and pending requests from previous slots. To avoid training multiple models with different input dimensions, we adopt a fixed-size concatenated state matrix. Zero-padding is applied when the number of available candidate paths is insufficient in paths matrix. However, excessive zero-padding may introduce unnecessary noise, especially when the actual number of candidate paths is much smaller than the predefined state matrix size.

The maximum number of concurrently processed requests is determined based on the network scale and connectivity. For the $10$-node network, the state matrix dimension is set to $10$. Given that the average number of arriving requests is $3$ and each request is assigned up to $k=2$ candidate paths, this configuration reduces redundant padding while maintaining a fixed input size for the learning model.

\subsubsection{Action} The model is trained to sequentially select paths. At each step $t$, the PPO agent selects an action $a_t$, corresponding to a path, according to the policy $\pi(a \mid s_t)$, where $s_t$ denotes the current state. The selected action is given by
\begin{equation}
\label{eqn:action}
a_t = \arg\max_{a} \, \pi(a \mid s_t).
\end{equation}

Once a path is selected, it is marked as unavailable. The agent continues this process until either all paths have been selected or no additional feasible paths can be accommodated. The final output is a set of parallelized paths.

\subsubsection{Reward} The reward $R$ for each parallel path group is designed to jointly encourage 
(i) efficient utilization of network resources and 
(ii) a high success rate of selected entanglement requests, while penalizing request failures. 
It is defined as
\begin{equation}
\label{eqn:reward}
R = \alpha R_{l} + \beta R_{s} - \gamma N_{f} .
\end{equation}

The link-level efficiency term is defined as
\begin{equation}
R_{l} = \frac{l_{s}}{l_{u}},
\end{equation}
where $l_{u}$ denotes the number of physical links utilized in the current time slot, and $l_{s}$ represents the subset of those links that successfully contribute to entanglement establishment. If multiple paths corresponding to a single request succeed, only the links associated with the path having the shortest execution time are counted in $l_{s}$. This ratio quantifies the efficiency with which allocated links lead to successful entanglement establishment between end nodes.

The request-level success ratio is defined as
\begin{equation}
R_{s} = \frac{N_{s}}{\min(N_{r}, N_{m})},
\end{equation}
where $N_{s}$ is the number of successfully completed requests, each having at least one successful entanglement path. $N_{r}$ represents the total number of requests in the current time slot, including newly arrived and pending requests from previous time slots, while $N_{m}$ is the maximum allowable number of requests, introduced to stabilize the learning process and mitigate reward inflation under heavy traffic conditions.

Furthermore, $N_{f}$ denotes the number of failed requests, introducing a penalty term that discourages inefficient or unstable routing decisions.

The coefficients $\alpha$, $\beta$, and $\gamma$ control the relative importance of each component and should be selected based on the training process as well as network conditions and topology. Typically, $\beta > \alpha$ biases the model toward improving overall request success rather than solely maximizing link efficiency, while a $\gamma$ penalizes failures of the selected requests.

\subsubsection{Loss Function}
The total loss $\mathcal{L}$ is a weighted sum of three terms: 
the clipped policy loss $\mathcal{L}_{\pi}$, the value loss $\mathcal{L}_{v}$, and an entropy bonus $\mathcal{H}$, as introduced in ref. \cite{schulman2017proximal}:
\begin{equation}
\mathcal{L} = \mathcal{L}_{\pi} + c_v \, \mathcal{L}_{v} - c_e \, \mathcal{H}
\end{equation}
where $c_v$ and $c_e$ are the value and entropy coefficients, respectively.

\subsubsection{Training}

The training dataset is generated from the same network topologies used in the simulations, allowing the PPO model to learn both the topology characteristics and quantum link properties. For each topology, the lowest-cost path information, including execution time, fidelity, and success status, is collected from the quantum simulator. These data are used exclusively during PPO training. During the evaluation of all allocation methods, identical request sequences are provided to ensure a fair comparison, and the allocated path information is independently generated by the simulator during evaluation for each method.

During training, multiple runs with different random seeds are performed to account for the stochastically introduced by varying request arrivals, including the number and types of requests across different time slots.

The PPO model employs a shared feature encoder and separate actor and critic networks. The shared encoder consists of two fully connected layers with $256$ neurons each, followed by ReLU activation functions to extract feature representations from the input state.

The actor network uses a single fully connected output layer with $256$ neurons to generate one logit for each candidate path, which is converted into action probabilities using a softmax function. The critic network consists of two fully connected layers with $256$ hidden neurons and a ReLU activation function. Mean pooling is applied to the shared feature representations before the critic estimates the state value.

To improve training stability, the PPO model is updated every $1024$ steps with a batch size of $64$ for $3$ optimization iterations. A cosine scheduler is employed to gradually reduce the learning rate from $1\times10^{-4}$ to $1\times10^{-7}$. A larger update interval (e.g., $2048$ steps) provides more stable updates but slows reward convergence; therefore, $1024$ steps is selected as a suitable trade-off for the considered quantum network simulation dataset. Conversely, an excessively large learning rate may lead to unstable policy updates due to the varying number and types of requests across time slots.

Other PPO parameters include a discount factor of $0.99$ and a smoothing parameter of $0.95$ for Generalized Advantage Estimation (GAE). The policy clipping ratio is set to $0.1$, with a value coefficient $c_v=0.15$ and an entropy regularization coefficient $c_e=0.01$.

The selection of the reward weights $\alpha$ and $\beta$ affects the learning performance. When most requests can be successfully served, the success ratio $R_s$ approaches $1$ in most time slots, causing the term $\beta R_s$ to become nearly constant. In this case, improving link efficiency through the $\alpha$ term becomes more important. However, in general scenarios, $\beta$ should be larger than $\alpha$ to prioritize maximizing the overall request success ratio.

The network connectivity and size also affect PPO learning performance. In low-connectivity networks, the limited availability of feasible paths restricts exploration and reduces optimization opportunities. Conversely, in highly connected networks, entanglement establishment becomes relatively easy, causing $R_s$ to approach $1$. In the meantime, the potential improvement in link efficiency diminishes because the available paths are already near-optimal with fewer connecting hops. Therefore, networks with moderate connectivity offer a favourable balance between path diversity and optimization potential, enabling more effective PPO learning.

The network size also affects convergence speed. For the medium-size quantum network considered in this paper ($20$ nodes), the reward converges after approximately $1000$ training epochs. For a larger network with $30$ nodes, convergence requires approximately $1500$ epochs, whereas a smaller 10-node network converges in fewer than $1000$ epochs. This increase in training time is mainly due to the larger number of possible path combinations and more complex network states that the PPO model must explore.

The training configurations discussed above are summarized in Table~\ref{tab:training_configuration}.

\begin{table}
\caption{\label{tab:training_configuration}Parameters used for PPO model training.}
\setlength{\tabcolsep}{3pt}
\renewcommand{\arraystretch}{1.2}
\centering
\begin{tabular}{|p{120pt}|p{108pt}|}
\hline
\textbf{Description} & \textbf{Value} \\
\hline
Learning rate & 
Decay from $1\times10^{-4}$ to $1\times10^{-7}$ \\
\hline
Update frequency & 
Every $1024$ steps \\
\hline
Batch size & 
$64$ \\
\hline
Number of optimization iterations &
$3$ \\
\hline
Value coefficient $c_v$ &
$0.15$ \\
\hline
Entropy regularization coefficient $c_e$ &
$0.01$ \\
\hline
Link efficiency coefficient $\alpha$ &
$0.4$ \\
\hline
Success ratio coefficient $\beta$ &
$0.6$ \\
\hline
Failure requests coefficient $\gamma$ &
$0.01$ \\
\hline
\end{tabular}
\end{table}

\subsection{Benchmark}
To serve as benchmarks for the proposed resource allocation methods described above, two allocation methods are used, which are based on the First-In-First-Out (FIFO) strategy.

The FIFO allocation strategy assigns entanglement resources strictly based on the order of request arrival, without prioritization to optimize objectives such as maximizing the number of accommodated requests or improving success rates. Requests that cannot be served within the current time slot, due to insufficient physical channels or quantum memories, are deferred and granted priority over newly arrived requests in the subsequent time slot.

\subsubsection{Dynamic FIFO Allocation}

In the Dynamic FIFO allocation, similar to the Dynamic Efficient approach, the lowest cost path for subsequent requests is updated dynamically after each request is served. This enables later requests to better accommodate previously selected requests.

\subsubsection{Static FIFO Allocation}
In contrast, the Static FIFO allocation method assigns each request a fixed path corresponding to the lowest cost based on the initial network topology. Consequently, subsequent requests cannot exploit dynamic adjustments to alternative paths, but each request is guaranteed its lowest cost path, similar to the Static Efficient method.

\subsection{Computational complexity Analysis}

The computational cost of the allocation methods determines how frequently scheduling decisions can be updated as the network size and request load increase. 
Let $\mathcal{R}$ denote the number of requests considered in a time slot, $V$ and $E$ the sets of nodes and physical links, respectively, and $k$ the maximum number of candidate paths generated per request. 

Dynamic Efficient and Dynamic FIFO recalculate the lowest-cost paths of the remaining requests after each allocation and graph update. The number of path calculations is therefore quadratic in the number of requests, giving a scheduling complexity of $O(\mathcal{R}^2(|E|+|V|\log|V|))$. 
After each request allocation, the available sub-graph is updated by reducing the capacities of the allocated links and removing unavailable links. Consequently, both the number of remaining requests and the size of the sub-graph decrease during the allocation process.
The quadratic dependence on $\mathcal{R}$ results from the iterative reconstruction of the available sub-graph and the subsequent lowest-cost path recalculation for the remaining unallocated requests after each request allocation. Since each lowest-cost path computation requires $O(|E|+|V|\log|V|)$ time, and the algorithm performs
$\sum_{r=1}^{\mathcal{R}} r = O(\mathcal{R}^{2})$
such computations throughout the allocation process, the overall computational complexity is
$O(\mathcal{R}^{2}(|E|+|V|\log|V|))$. 
In contrast, Static Efficient and Static FIFO calculate the lowest-cost path only once for each request. Their complexity is $O(\mathcal{R}(|E|+|V|\log|V|))$.

The computational complexity of the Success Enhancement allocation method is $O\!\left(\mathcal{R}k|V|\left(|E|+|V|\log|V|\right)\right)$, where the term $O\!\left(k|V|\left(|E|+|V|\log|V|\right)\right)$ corresponds to the complexity of generating the $k$ lowest-cost paths for each request.The complexity is dominated by the generation of the $k$ lowest-cost candidate paths.

The computational complexity of the MILP-based allocation method consists of two components: candidate path generation and integer optimization. The candidate path generation requires $O\!\left(\mathcal{R}k|V|\left(|E|+|V|\log|V|\right)\right)$ operations. The MILP formulation introduces $O(\mathcal{R}k)$ binary decision variables along with corresponding conflict constraints. As binary integer programming is NP-hard, the worst-case computational complexity of solving the MILP is exponential in the number of binary variables. Therefore, for a fixed number of candidate paths $k$, the complexity is bounded by $O(2^{\mathcal{R}k})$.

For PPO, the training is performed offline and is excluded from the online scheduling complexity.The online scheduling cost consists of candidate-path generation and a sequence of actor–critic forward passes.
Let $n\leq \mathcal{R}k$ be the number of candidate paths, $A$ the number of sequential path-selection steps, $I_d$ the input-feature dimension, and $h$ the hidden-layer width. The candidate path generation requires $O(\mathcal{R}P_k)$ operations. The complexity of one actor–critic forward pass is $C_{\mathrm{NN}}(n)=O\left(n(I_dh+h^2)+h^2\right)$. Because PPO selects paths sequentially, its total online complexity is is $O(\mathcal{R}k|V|\left(|E|+|V|\log|V|\right)+AC_{\mathrm{NN}}(n))$. For a fixed neural-network architecture, $C_{\mathrm{NN}}$ scales approximately linearly with $n$. Since $A$ can also increase linearly with $n$, the worst-case PPO inference complexity is quadratic in the number of candidate paths. Training each PPO model required approximately one hour using a NVIDIA RTX A4000 GPU. Although separate training may be required for substantially different network configurations, this offline training time remains manageable and does not contribute to online scheduling latency.

Computational complexities and average running time of the allocation methods are summarized in Table~\ref{tab:computational_complexity_Analyse}. The average running time is calculated by averaging the running times recorded for each time slot across all methods. The simulations are conducted for a 30-node quantum network using an AMD Ryzen Threadripper PRO 5995WX 64-core processor with 128~GiB of system memory. The computation times are given here to provide a practical comparison of the scheduling overhead of the methods for the evaluated scenario, complementing the theoretical complexity analysis. These values are not intended as universal deployment latencies, since they depend on the software implementation and processing hardware.

Overall, Static Efficient and Static FIFO have the lowest complexity because paths are calculated only once for each request.
This is followed by Success Enhancement that introduces additional candidate-path evaluation but remains polynomial, particularly when the number of candidate paths is small. 
Dynamic Efficient and Dynamic FIFO methods have higher complexity because paths are recalculated following each allocation.
PPO also has polynomial online complexity, consisting of path calculation and repeated actor–critic inference. Its computational requirement is higher than that of the static methods and depends on the number of selection steps and candidate paths, but it avoids the exponential worst-case complexity of MILP. 
MILP complexity grows exponentially with the number of candidate-path variables, making it the least scalable method, although it can provide an optimization benchmark.

\begin{table}
\caption{\label{tab:computational_complexity_Analyse}Computational complexities and average computation times of the allocation methods.}
\setlength{\tabcolsep}{3pt}
\renewcommand{\arraystretch}{1.2}
\centering
\begin{tabular}{|p{70pt}|p{110pt}|p{40pt}|}
\hline
\textbf{Allocation methods} & \textbf{Computational complexity} & \textbf{Avg. comp. time (ms)} \\
\hline
Dynamic Efficient &
$O(\mathcal{R}^2(|E|+|V|\log|V|))$ &
3.43 \\
\hline
Dynamic FIFO &
$O(\mathcal{R}^2(|E|+|V|\log|V|))$ &
3.29 \\
\hline
Static Efficient &
$O(\mathcal{R}(|E|+|V|\log|V|))$ &
0.66 \\
\hline
Static FIFO &
$O(\mathcal{R}(|E|+|V|\log|V|))$ &
0.73 \\
\hline
Success Enhancement &
$O\!\left(\mathcal{R}k|V|\left(|E|+|V|\log|V|\right)\right)$ &
3.77 \\
\hline
MILP &
$O(2^{\mathcal{R}k})$ &
166.62 \\
\hline
PPO &
$O(\mathcal{R}k|V|\left(|E|+|V|\log|V|\right)+AC_{\mathrm{NN}}(n))$, excluding offline training &
18.39 \\
\hline
\end{tabular}
\end{table}

\section{Performance Evaluation}

In this section, we present the simulation settings and evaluate the performance of the proposed allocation methods in terms of the delay time, number of successful requests, capacity utilization, and request handling rate, as defined in Section \ref{sec: prob_formulation}.

\label{sec: evaluation}
\subsection{Simulation settings}

This paper targets near-term metropolitan quantum networks, where the network size is typically expected to be in the order of tens of nodes. The evaluation of $10$, $20$, and $30$ nodes networks represents small, medium, and relatively large scale scenarios within this range. 

To investigate the performance of resource allocation methods in different types of network topologies, two classes of random graphs were generated. The first is the Watts-Strogatz graph $G(N,K,r_p)$ \cite{watts1998collective}. Each graph is initialized as a regular ring lattice with $N$ nodes and $K$ edges per node, after which each edge is independently rewired with probability $r_p$. This construction yields networks with high clustering. A higher rewiring probability results in shorter path lengths and shortcuts connecting distant nodes.

The second is a random geometric graph $G(N,r_e)$, in which connectivity is constrained by physical distance. In this topology, $N$ nodes are distributed uniformly at random within a bounded region, and an undirected edge is established between any pair of nodes separated by a Euclidean distance that does not exceed $r_e$. The radius $r_e$ in unit square controls the network density: larger values of $r_e$ increases the average number of links per node and reduce path lengths, while smaller values can produce sparse or disconnected networks.

The primary difference between these two topologies lies in node distribution. In the Watts-Strogatz graph, nodes are initially arranged in a ring, forming a single clustered structure. In contrast, in the random geometric graph, nodes are randomly distributed within a unit square, forming multiple local clusters determined by the radius $r_e$. As $r_e$ increases, the graph becomes more connected; however, it does not form a single cluster as in the Watts-Strogatz graph.

To ensure comparable hop distances, we preserve the node connection within the generated topology while assigning hop distances randomly from the set $\{5,6,7,8,9,10,11,12,13,14,15\}$~km. Moreover, the fiber attenuation rate in Eq.~(\ref{eq:photon_loss}) is randomly sampled from the predefined set in Table~\ref{tab:quatum_parameters}. Once assigned, all hops characteristics remain fixed.

Each allocation method was evaluated over $1000$ time slots, with a duration of $80{,}000$~ns per slot. This duration was chosen to ensure that most paths have sufficient time to complete entanglement generation between their end nodes, based on the simulation results of the node chain entanglement.

\begin{figure*}
  \centering

  \begin{subfigure}{0.47\textwidth}
    \centering
    \includegraphics[width=\linewidth]{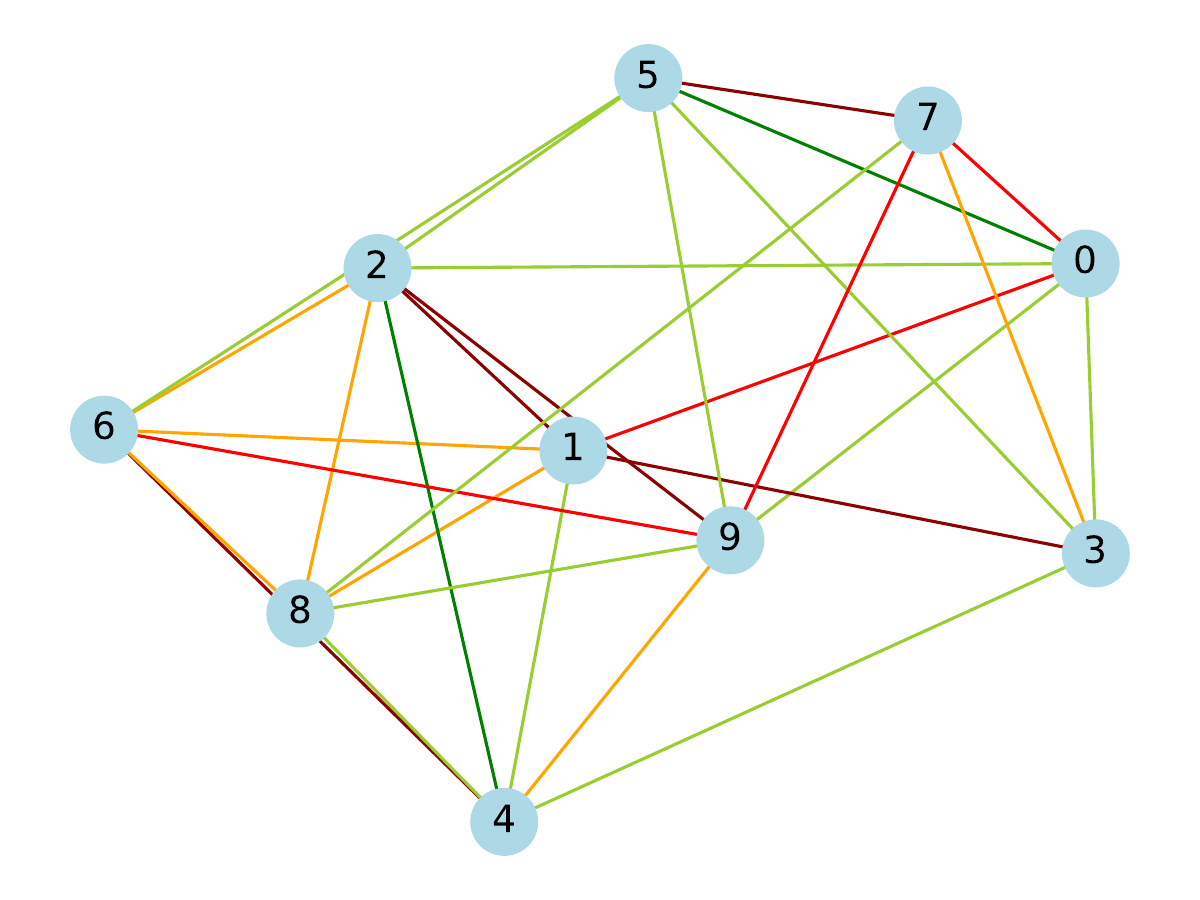}
    \caption{Small network with rewire probability $r_p = 0.25$.}
    \label{fig:s_0.25_network}
  \end{subfigure}
  \hfill
  \begin{subfigure}{0.47\textwidth}
    \centering
    \includegraphics[width=\linewidth]{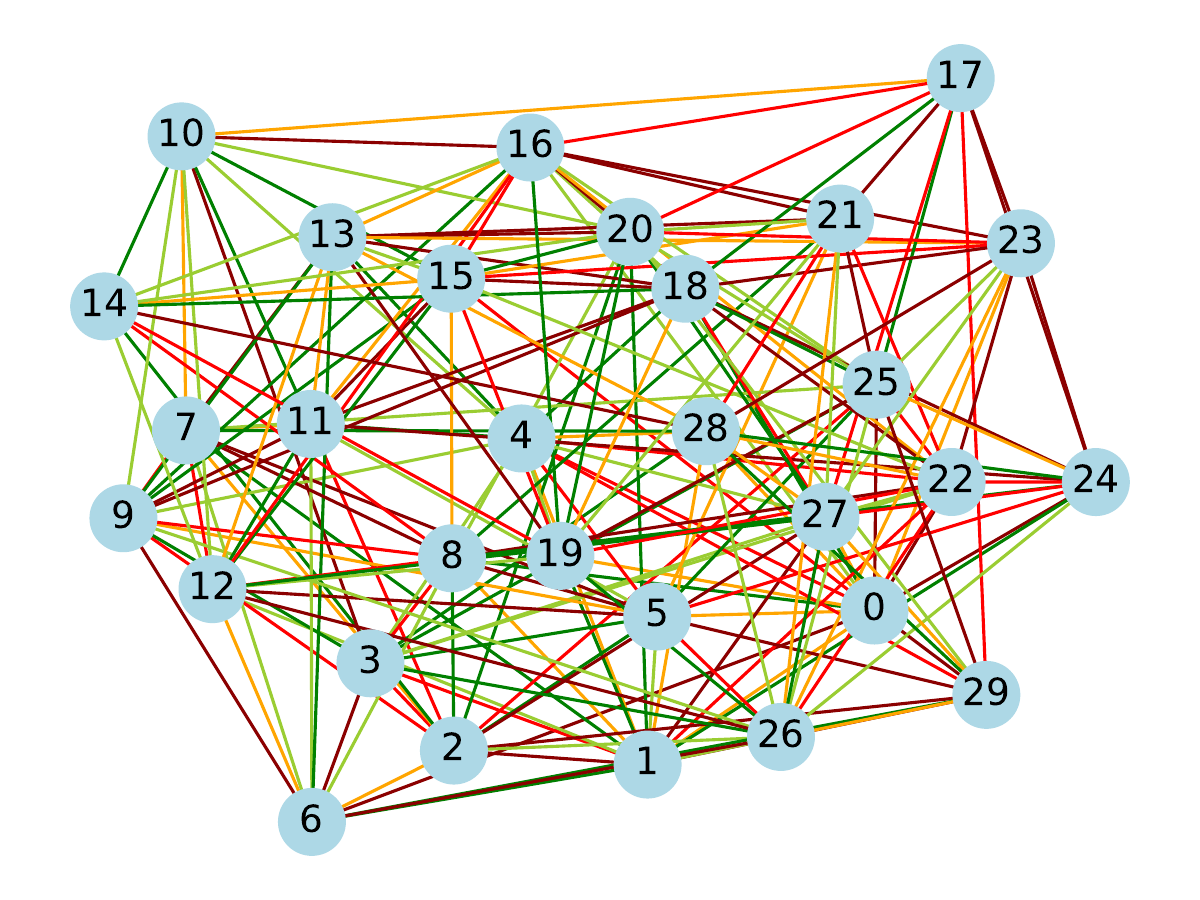}
    \caption{Large network with rewire probability $r_p = 0.25$.}
    \label{fig:l_0.25_network}
  \end{subfigure}

  \vspace{0.5em}

  \begin{subfigure}{0.47\textwidth}
    \centering
    \includegraphics[width=\linewidth]{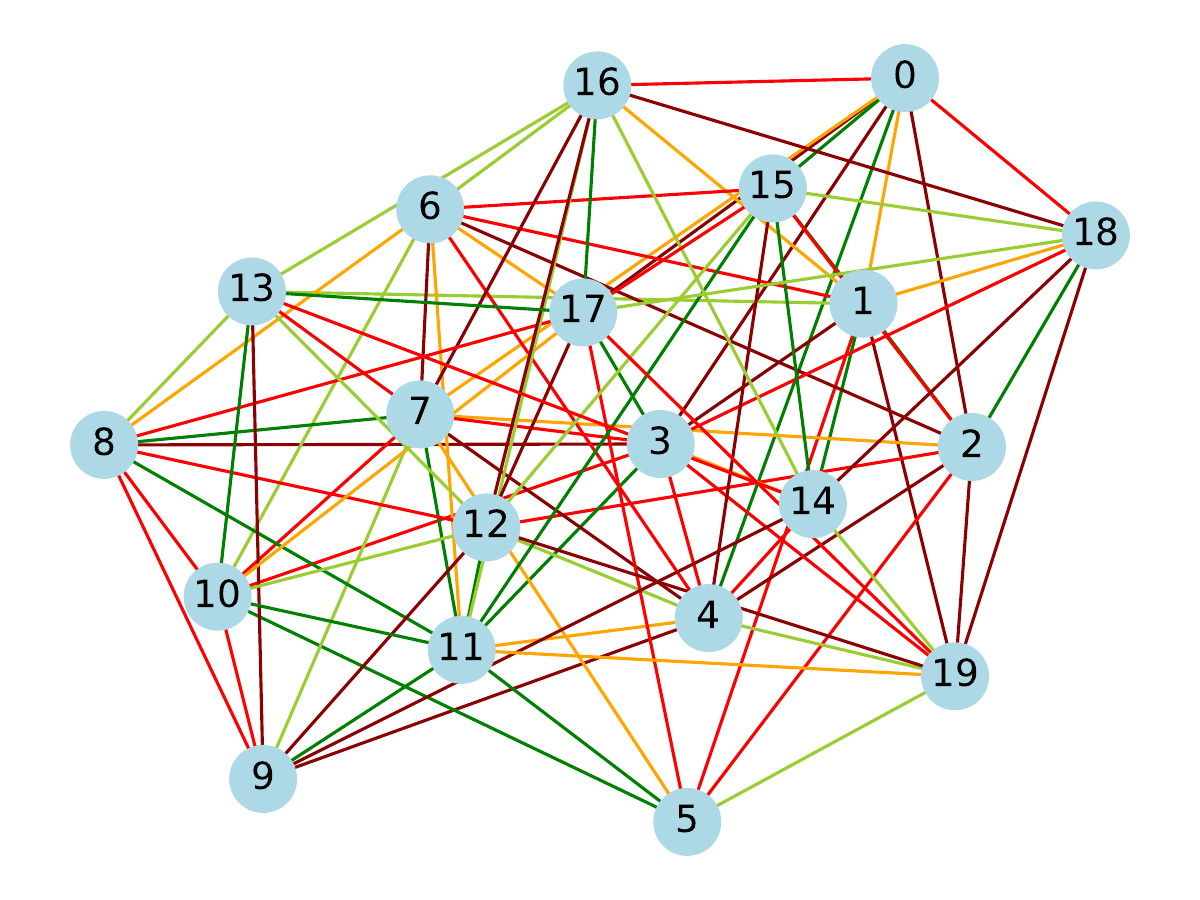}
    \caption{Watts-Strogatz network with rewire probability $r_p = 0.25$.}
    \label{fig:ws_network}
  \end{subfigure}
  \hfill
  \begin{subfigure}{0.47\textwidth}
    \centering
    \includegraphics[width=\linewidth]{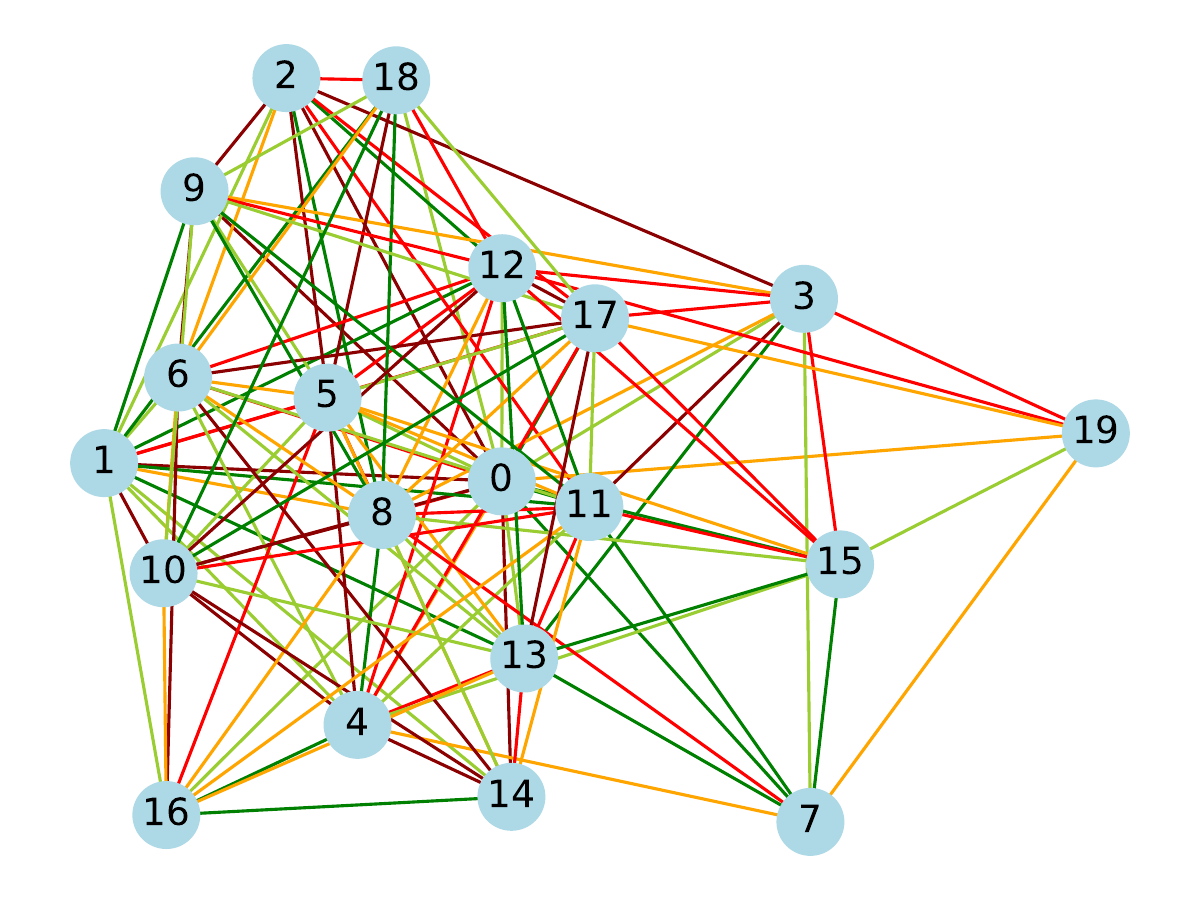}
    \caption{Random Geometric network $r_e =0.5$.}
    \label{fig:rgg_network}
  \end{subfigure}

  \caption{Network topologies of varying sizes and types. Links are colour-coded by photon loss rate [dB/km]: green ($0.15$), dark green ($0.20$), orange ($0.25$), red ($0.30$), and dark red ($0.35$). Warmer colours indicate higher photon loss rate.}
  \label{fig:network_topology}
  \hrulefill
\end{figure*}

In summary, both network size and connectivity were varied in the simulation, as illustrated in Fig.~\ref{fig:network_topology}. The parameters used for analysing network size variations are summarized in Table~\ref{tab2}. For the connectivity analysis, topologies generated using the Watts-Strogatz and random geometric graph methods are compared on a $20$-node network with network load $\lambda = 6$ and a maximum of $8$ arrivals per time slot. Given the limited resource availability in the quantum network, the simulations use a moderate level of load that creates meaningful resource contention without causing persistent queue growth, thereby enabling the allocation methods to be compared under stable and consistent operating conditions. Other relevant parameters are summarized in Table~\ref{tab3}.

\begin{table*}
\centering
\caption{Simulation parameters for small, medium, and large network sizes.}
\label{tab:simulation_parameters}
\setlength{\tabcolsep}{6pt}
\renewcommand{\arraystretch}{1.2}
\begin{tabular}{|p{155pt}|p{80pt}|p{80pt}|p{80pt}|}
\hline
\textbf{Description} & \textbf{Small Network} & \textbf{Medium Network} & \textbf{Large Network} \\
\hline
Number of nodes $N$ & 10 & 20 & 30 \\
\hline
Average edges per
vertex $K$ & 6 & 10 & 14 \\
\hline
Number of links $L$ & 30 & 100 & 210\\
\hline
Network load $\lambda$ & 2 & 6 & 8 \\
\hline
Max number of arrival requests per time slot $N_m$ & 4 & 8 & 14 \\
\hline
\end{tabular}
\label{tab2}
\end{table*}

\begin{table*}
\centering
\caption{Simulation parameters for different topologies in the medium-size network.}
\label{tab:simulation_parameters}
\setlength{\tabcolsep}{6pt}
\renewcommand{\arraystretch}{1.2}
\begin{tabular}{|p{150pt}|p{130pt}|p{130pt}|}
\hline
\textbf{Description} & \textbf{Watts-Strogatz Graph}  & \textbf{Random Geometric Graph} \\
\hline
Topology parameters & 
$K = 10$, $r_p = 0.25$ 
&$r_e = 0.5$ \\
\hline
Number of links $L$ & 100 & 132 \\
\hline
\end{tabular}
\label{tab3}
\end{table*}

\subsection{Results}

Resource allocation methods and their benchmarks are evaluated across varying network sizes, retry mechanisms, and different topology types. 

\subsubsection{Varied network size}

The network size is varied from $10$ to $30$ nodes (Figs. \ref{fig:s_0.25_network}-\ref{fig:l_0.25_network}) while maintaining a similar level of connectivity, and no retry mechanism is performed for entanglement requests. The network load is adjusted according to the network size, which are summarized in Table~\ref{tab2}.

\begin{figure*}
  \centering
  \begin{subfigure}{0.33\textwidth}
    \centering
    \includegraphics[width=\linewidth]{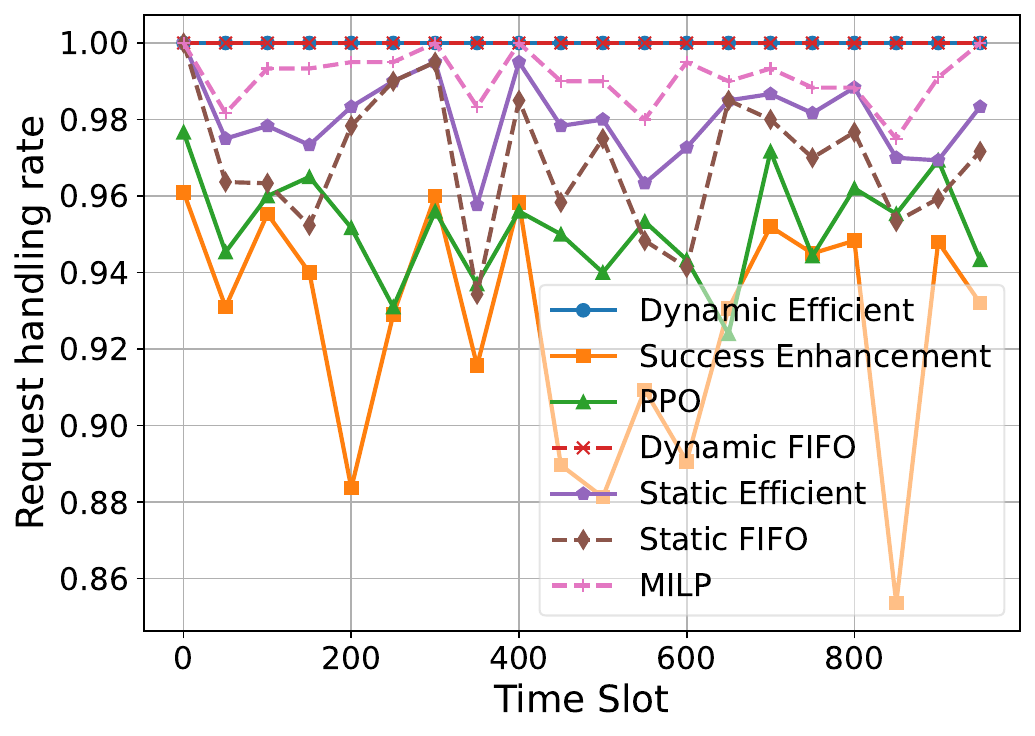}
    \caption{Small network.}
  \end{subfigure}
  \hfill
  \begin{subfigure}{0.33\textwidth}
    \centering
    \includegraphics[width=\linewidth]{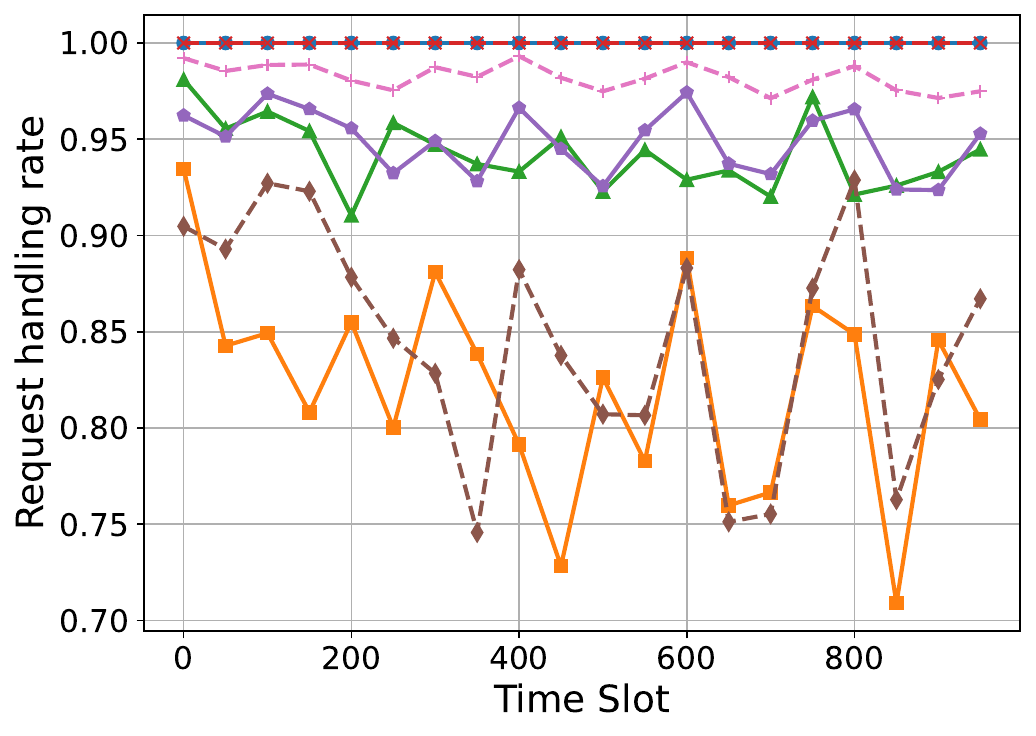}
    \caption{Medium network.}
  \end{subfigure}
  \hfill
  \begin{subfigure}{0.33\textwidth}
    \centering
    \includegraphics[width=\linewidth]{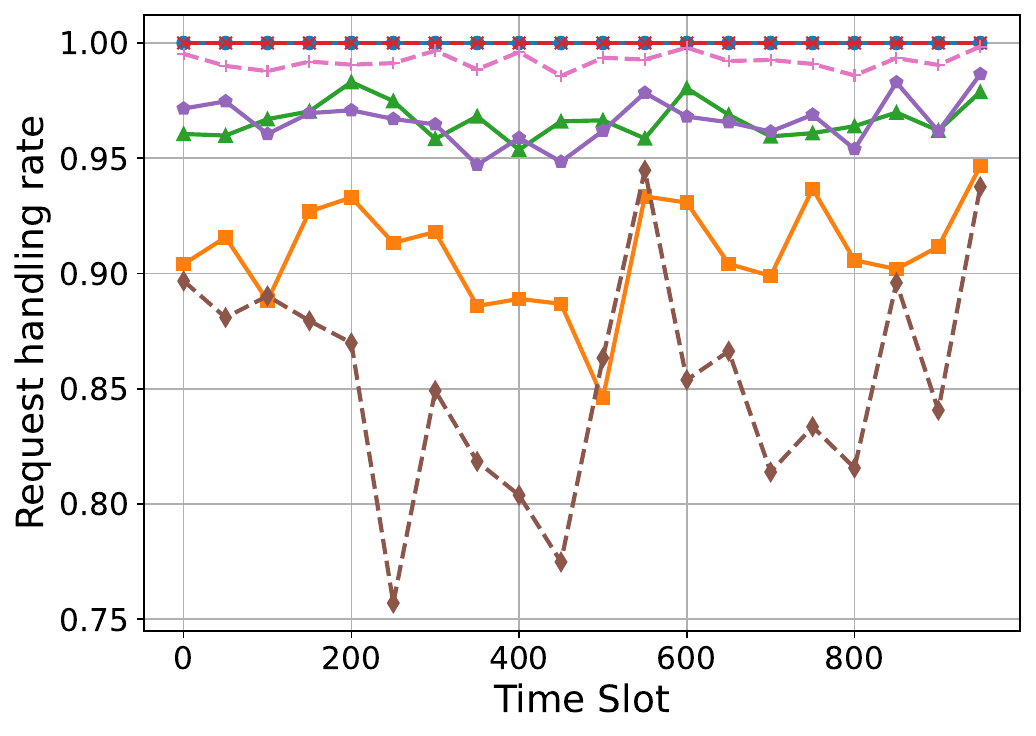}
    \caption{Large network.}
  \end{subfigure}
  \caption{Average request handling rate over $50$ time slots with different network size.} 
  \label{fig:handle_rate_compare}
  \hrulefill
\end{figure*}

The Dynamic Efficient and Dynamic FIFO algorithms achieve relatively low mean delays under varying network loads, as shown in Table~\ref{tab:varied_network_size_mean_delay}. However, they consistently have lower number of success entanglement requests than the other methods, as shown in Table~\ref{tab:varied_network_size_num_success}. This is because these approaches allocate any available path to pending requests, resulting in a consistently high request handling rate close to $1$ in Fig.~\ref{fig:handle_rate_compare}. Consequently, successfully executed requests have lower delays, as they are rarely deferred to subsequent time slots. Nevertheless, this strategy often forces subsequent requests to use higher-cost paths, causing them to fail the fidelity requirement.

As the number of incoming requests increases and the handling rate decreases, both methods experience higher congestion, reducing the advantages of optional path selection in other multi-path allocation schemes. Under heavy traffic conditions, Dynamic Efficient Allocation outperforms Dynamic FIFO in terms of average delay, owing to its more efficient utilization of constrained network resources, as also demonstrated by the corresponding static schemes.

For both the Static Efficient and Static FIFO methods, the lowest-cost path is fixed and repeatedly used for all requests. This design increases the likelihood of success, as requests consistently traverse the most reliable path. However, it results in a lower handling rate compared with their dynamic versions, as shown in Fig.~\ref{fig:handle_rate_compare}. Moreover, due to the strict request handling order, the Static FIFO method exhibits the worst delay performance as the network size varies.

Regarding the number of successful requests, both the Success Enhancement strategy, MILP and the PPO-based allocation approach achieve the highest success counts. The reason these methods complete more entanglement requests is that they provide multiple candidate paths for each request.

\begin{figure*}
  \centering
  \begin{subfigure}{0.33\textwidth}
    \centering
    \includegraphics[width=\linewidth]{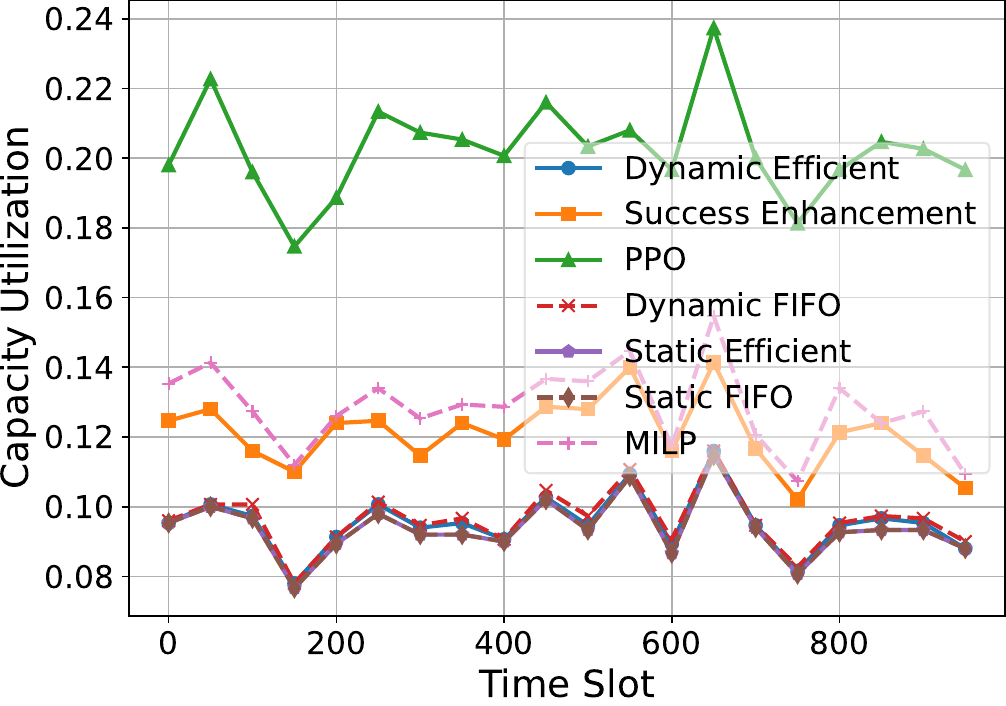}
    \caption{Small network.}
  \end{subfigure}
  \hfill
  \begin{subfigure}{0.33\textwidth}
    \centering
    \includegraphics[width=\linewidth]{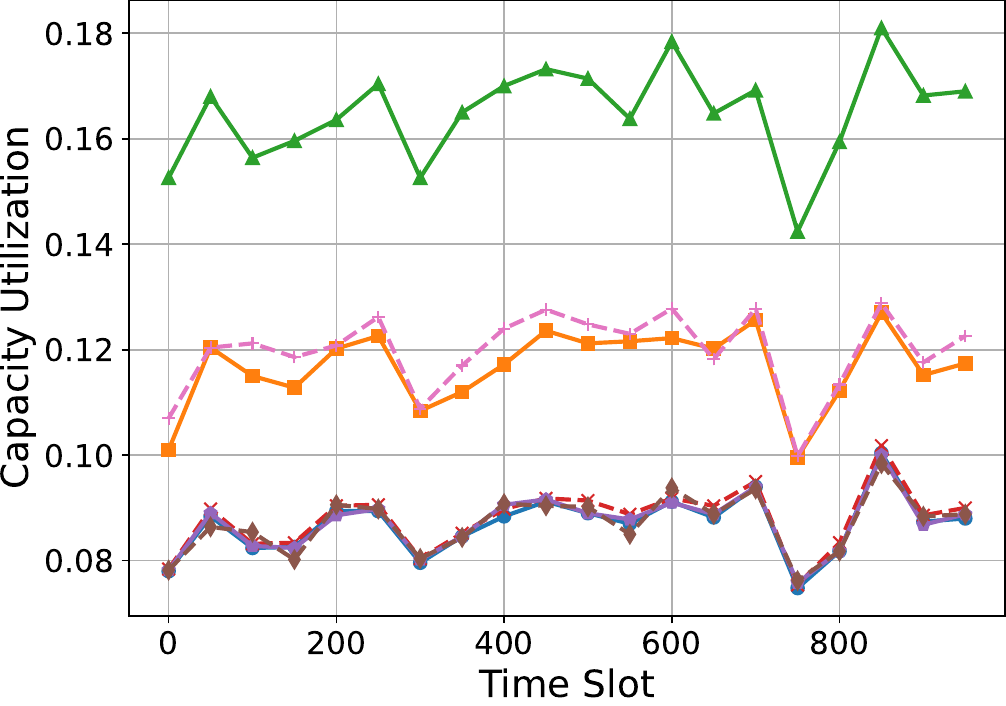}
    \caption{Medium network.}
  \end{subfigure}
  \hfill
  \begin{subfigure}{0.33\textwidth}
    \centering
    \includegraphics[width=\linewidth]{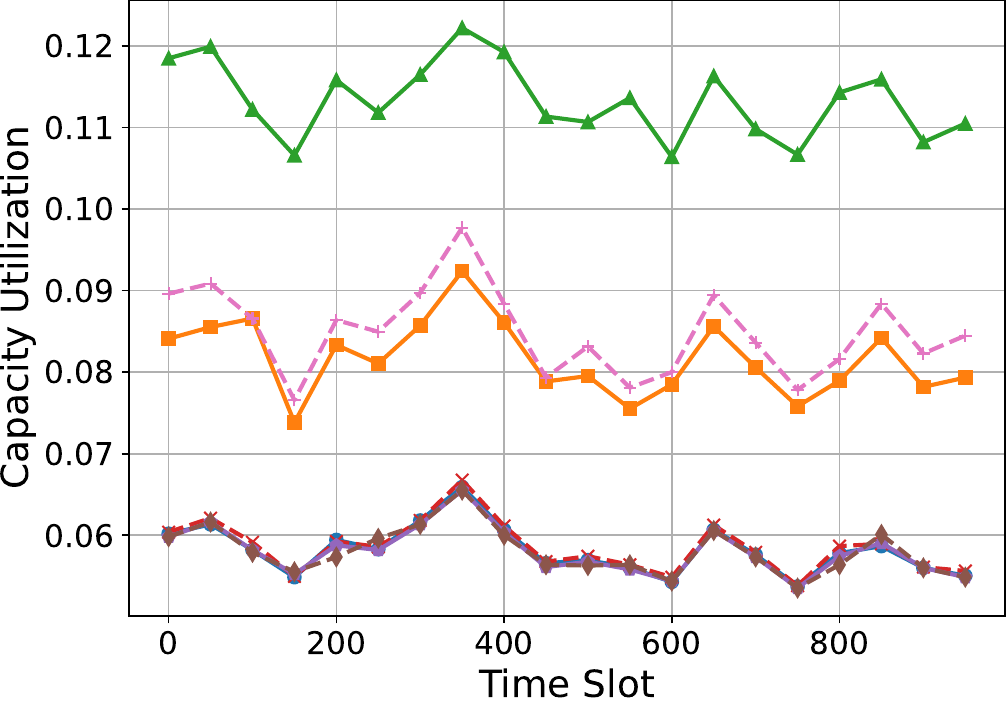}
    \caption{Large network.}
  \end{subfigure}
  \caption{Average capacity utilization over $50$ time slots with different network size.}
  \label{fig:capacity_compare}
  \hrulefill
\end{figure*}

The Success Enhancement strategy prioritizes medium-worst requests over other requests, which may increase the waiting time of queued requests. In contrast, the MILP approach optimally selects requests and their corresponding paths while satisfying edge-disjoint resource constraints, thereby achieving both a higher request handling rate and more path attempts. Similarly, the PPO model jointly optimizes request handling and success rate, at the cost of higher capacity utilization (as shown in Fig.~\ref{fig:capacity_compare}). Consequently, both the MILP and PPO approaches achieve consistently lower average delays per successful request than the Success Enhancement strategy.

\begin{table*}
\centering
\caption{Mean delay ($\mu$s) for different algorithms under varying network size}
\renewcommand{\arraystretch}{1.2}
\begin{tabular}{|p{2cm}|p{2.2cm}|p{2cm}|p{1.8cm}|p{1.8cm}|p{2.5cm}|p{1cm}|p{1cm}|}
\hline
Network size & Dynamic Efficient & Dynamic FIFO & Static Efficient & Static FIFO & Success Enhancement & PPO & MILP\\
\hline
Small(Fig. \ref{fig:s_0.25_network})
& 203.35 & 202.96 & 235.17 & 244.30 & 282.79 & 226.97 & 202.83\\
\hline
Medium(Fig. \ref{fig:ws_network})
& 179.38 & 180.15 & 221.45 & 384.45 & 336.41 & 205.79 & 177.22\\
\hline
Large(Fig. \ref{fig:l_0.25_network})
& 182.75 & 183.00 & 209.17 & 376.71 & 249.92 & 183.14 & 170.21\\
\hline
\end{tabular}
\label{tab:varied_network_size_mean_delay}
\end{table*}

\begin{table*}
\centering
\caption{Number of successes for different algorithms under varying network size}
\renewcommand{\arraystretch}{1.2}
\begin{tabular}{|p{2cm}|p{2.2cm}|p{2cm}|p{1.8cm}|p{1.8cm}|p{2.5cm}|p{1cm}|p{1cm}|}
\hline
Network size & Dynamic Efficient & Dynamic FIFO & Static Efficient & Static FIFO & Success Enhancement & PPO & MILP\\
\hline
Small(Fig. \ref{fig:s_0.25_network})
& 1914 & 1888 & 1938 & 1907 & 1979 & 1978 & 1979\\
\hline
Medium(Fig. \ref{fig:ws_network})
& 5204 & 5228 & 5301 & 5253 & 5408 & 5459 & 5429\\
\hline
Large(Fig. \ref{fig:l_0.25_network})
& 7190 & 7170 & 7256 & 7226 & 7420 & 7459 & 7441\\
\hline
\end{tabular}
\label{tab:varied_network_size_num_success}
\end{table*}

\subsubsection{Varied maximum entanglement retry times}

The $20$-node network (Fig. \ref{fig:ws_network}) is considered to investigate the impact of the maximum number of retries per request. As shown in Table~\ref{tab:max_tries_num_success}, when no retries are allowed, the advantage of multipath allocation is evident. PPO, MILP and Success Enhancement achieve the highest number of successful requests, outperforming methods that assign only a single path per entanglement request. 

\begin{figure*}
  \centering
  \begin{subfigure}{0.33\textwidth}
    \centering
    \includegraphics[width=\linewidth]{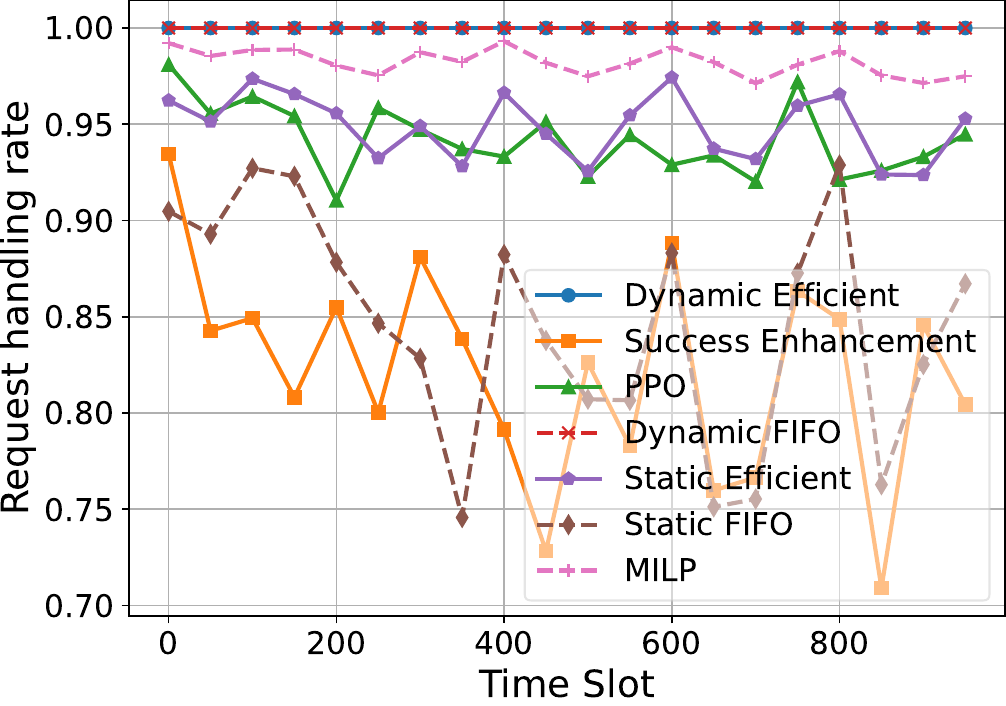}
    \caption{No retry.}
  \end{subfigure}
  \hfill
  \begin{subfigure}{0.33\textwidth}
    \centering
    \includegraphics[width=\linewidth]{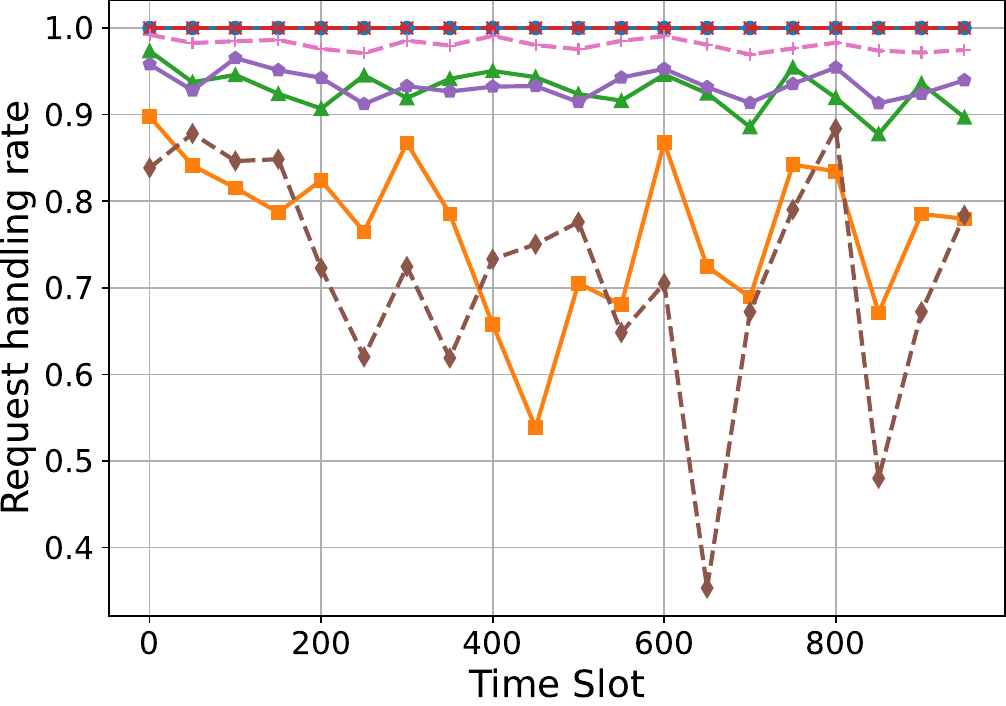}
    \caption{Maximum one retry.}
  \end{subfigure}
  \hfill
  \begin{subfigure}{0.33\textwidth}
    \centering
    \includegraphics[width=\linewidth]{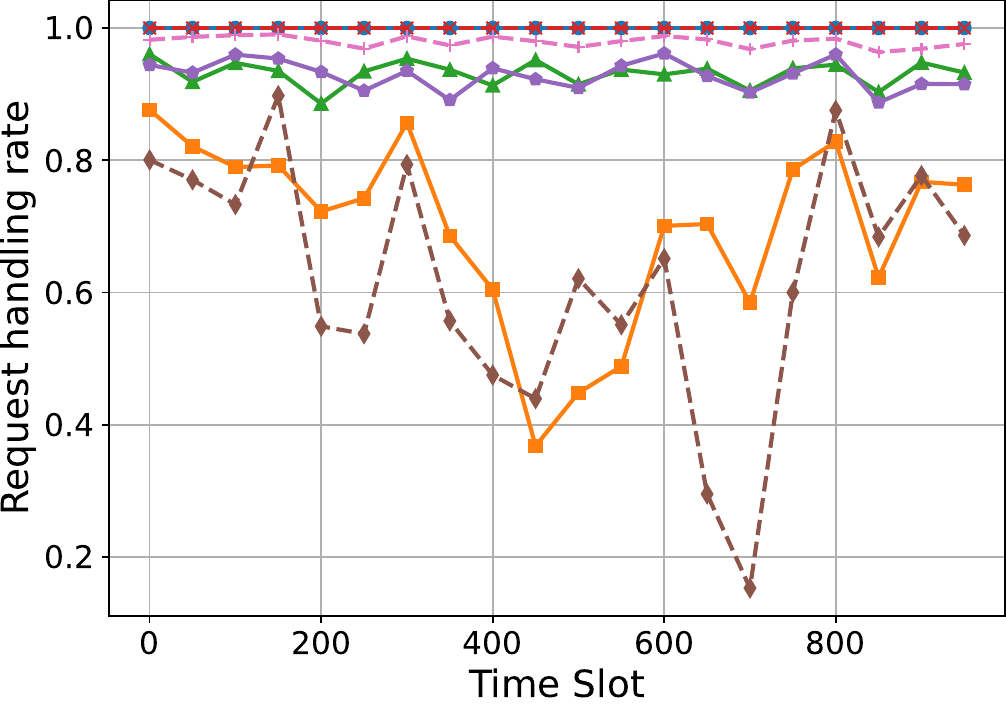}
    \caption{Maximum two retries.}
  \end{subfigure}
  \caption{Average request handling rate over $50$ time slots with maximum number of retries per entanglement request.} 
  \label{fig:handle_rate_tries_compare}
  \hrulefill
\end{figure*}

However, when retries are incorporated, allowing failed requests to be reattempted in subsequent time slots, the advantage of multipath allocation diminishes. With a maximum of $2$ retry attempts per request, the number of successful requests becomes comparable across all allocation methods. Under this retry mechanism, Dynamic Efficient and FIFO achieve higher success rates (Table~\ref{tab:max_tries_num_success}) while maintaining minimal delay (Table~\ref{tab:max_tries_mean_delay}), owing to their adaptive sub-graph updating strategy. They also attain the highest request handling rate (Fig.~\ref{fig:handle_rate_tries_compare}).

\begin{figure*}
  \centering
  \begin{subfigure}{0.33\textwidth}
    \centering
    \includegraphics[width=\linewidth]{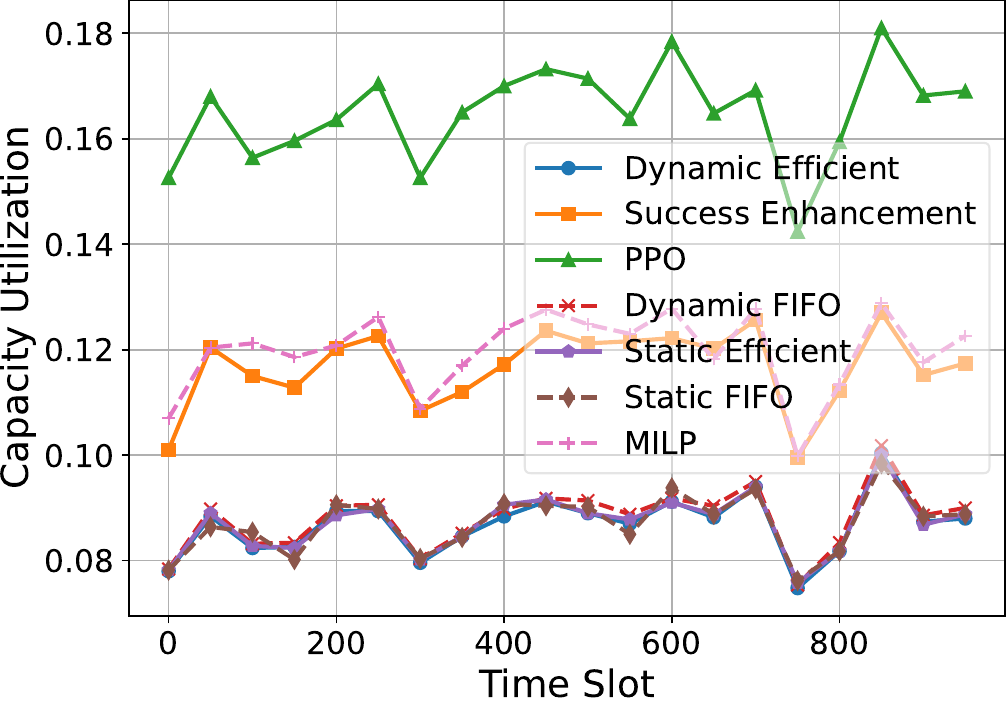}
    \caption{No retry.}
  \end{subfigure}
  \hfill
  \begin{subfigure}{0.33\textwidth}
    \centering
    \includegraphics[width=\linewidth]{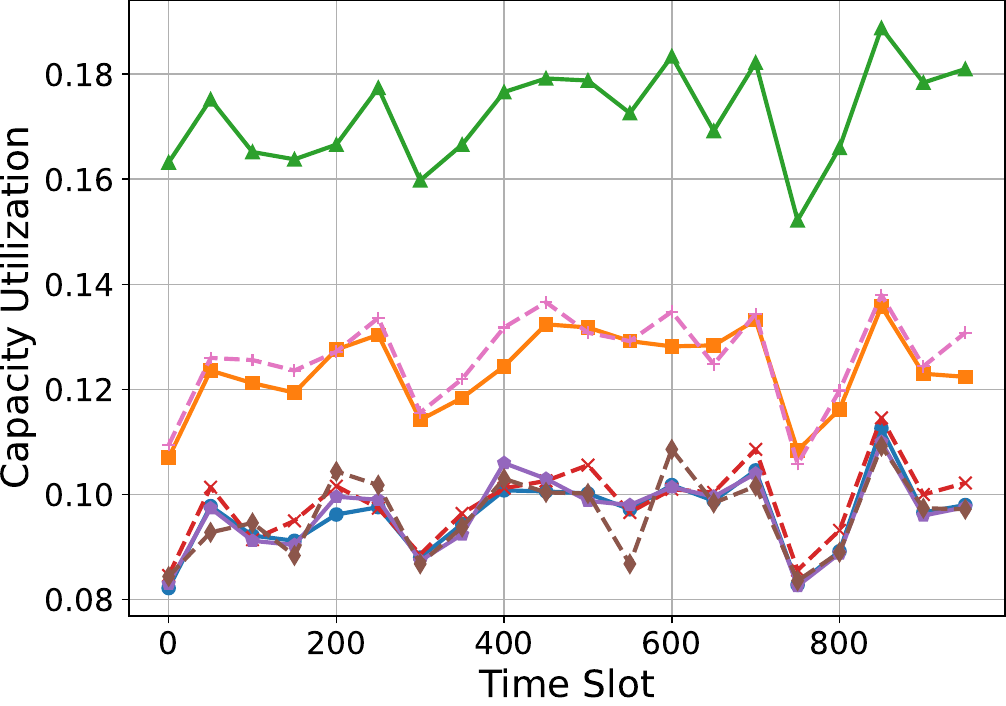}
    \caption{Maximum one retry.}
  \end{subfigure}
  \hfill
  \begin{subfigure}{0.33\textwidth}
    \centering
    \includegraphics[width=\linewidth]{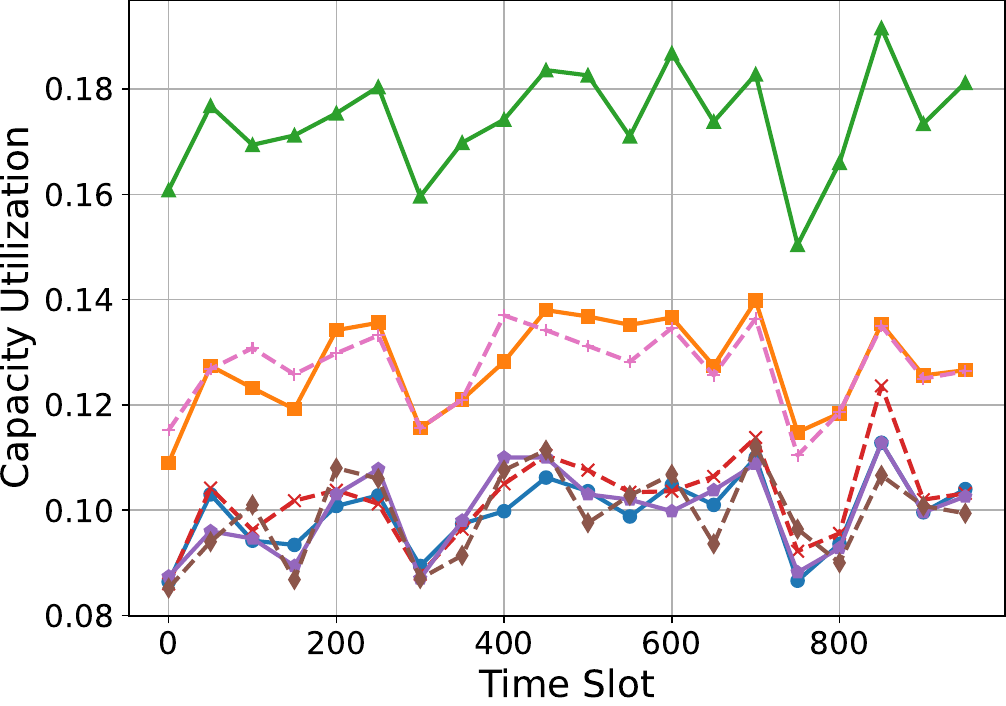}
    \caption{Maximum two retries.}
  \end{subfigure}
  \caption{Average capacity utilization over $50$ time slots with maximum number of retries per entanglement request.}
  \label{fig:capacity_tries_compare}
  \hrulefill
\end{figure*}

Compared with the Static Efficient approach, the PPO-based allocation method consistently achieves lower mean delay and a higher number of successfully served requests, at the expense of increased capacity utilization, as shown in Fig.~\ref{fig:capacity_tries_compare}, across different maximum retry limits. The MILP allocation method further achieves even lower mean delay than the dynamic approaches.

The Static FIFO achieves a comparable number of successful requests to Static Efficient across different retry times, as both preserve the lowest-cost path for each request. However, it exhibits the highest mean delay, even exceeding that of Success Enhancement, as it only adheres to the order of request arrivals and halts when the next request cannot be accommodated due to links taken up by previous selected requests.

\begin{table*}
\centering
\caption{Number of successes under different maximum entanglement retry times}
\renewcommand{\arraystretch}{1.2}
\begin{tabular}{|p{2cm}|p{2.2cm}|p{2cm}|p{1.8cm}|p{1.8cm}|p{2.5cm}|p{1cm}|p{1cm}|}
\hline
Maximum retries & Dynamic Efficient & Dynamic FIFO & Static Efficient & Static FIFO & Success Enhancement & PPO & MILP\\
\hline
0 & 5204 & 5228 & 5301 & 5253 & 5408 & 5459 & 5429\\
\hline
1 & 5616 & 5587 & 5577 & 5584 & 5607 & 5678 & 5674\\
\hline
2 & 5673 & 5635 & 5618 & 5613 & 5621 & 5712 & 5703\\
\hline
\end{tabular}
\label{tab:max_tries_num_success}
\end{table*}

\begin{table*}
\centering
\caption{Mean delay per request ($\mu$s) under different maximum entanglement retry times}
\renewcommand{\arraystretch}{1.2}
\begin{tabular}{|p{2cm}|p{2.2cm}|p{2cm}|p{1.8cm}|p{1.8cm}|p{2.5cm}|p{1cm}|p{1cm}|}
\hline
Maximum retries & Dynamic Efficient & Dynamic FIFO & Static Efficient & Static FIFO & Success Enhancement & PPO & MILP\\
\hline
0 & 179.38 & 180.15 & 221.45 & 384.45 & 336.41 & 205.79 & 177.22\\
\hline
1 & 236.95 & 238.23 & 276.29 & 816.64 & 399.44 & 263.10 & 212.13\\
\hline
2 & 255.35 & 250.10 & 288.64 & 1339.27 & 463.78 & 274.04 & 218.68\\
\hline
\end{tabular}
\label{tab:max_tries_mean_delay}
\end{table*}

\subsubsection{Varied network topology}

To analyse how the allocation method varies with network topology type, two medium-sized networks with a relatively medium number of links are considered (Fig.~\ref{fig:ws_network} and Fig.~\ref{fig:rgg_network}). To obtain a fair comparison, arriving requests per time slot is identical for both topologies.

The key structural difference is that the topology in Fig.~\ref{fig:ws_network} exhibits the clustering property of the Watts--Strogatz graph, resulting in higher local connectivity. In contrast, in the topology in Fig.~\ref{fig:rgg_network}, the cluster of nodes $3,7,15,19$ is relatively distant from the remaining nodes, which influences different delay performances among these allocation methods. 

Due to the relatively large distances between connected end nodes, Dynamic Efficient, Dynamic FIFO incur slightly higher delays in the Random Geometric Graph relative to the Watts--Strogatz graph, as shown in Table~\ref{tab:topology_mean_delay}. 

However, the delay in the Random Geometric Graph is lower under Static FIFO than in the Watts-Strogatz graph. This difference arises from the strict request-order handling of Static FIFO. When lowest path costs are comparable and the Random Geometric Graph contains more links than the Watts-Strogatz graph, the effect of request ordering is reduced, yielding a relatively lower delay.

The Static Efficient and Success Enhancement schemes exhibit lower delay and a greater number of successful entanglements in Random Geometric Graphs than in Watts--Strogatz graphs, primarily due to the higher number of links in Random Geometric Graphs.

The PPO allocation method performs well in both topologies, achieving the highest number of successful requests with relatively low delay, comparable to MILP allocation method. Success Enhancement also achieves a high success rate but exhibits the second-highest delay after Static FIFO.

In Table~\ref{tab:topology_num_success}, as the number of arriving requests per time slot is identical for both topologies, resulting in the same total number of requests throughout the simulation, the Random Geometric Graph achieves a higher number of successful entanglement requests than the Watts--Strogatz graph. This improvement is attributed to the higher number of links of the Random Geometric Graph, which provides more routing alternatives and increases the likelihood of selecting lower-cost paths, thereby improving entanglement generation success probability. Similarly, Table~\ref{tab:capacity_handle} shows that all methods achieve lower capacity utilization and higher request handling rates in the Random Geometric Graph topology compared with the Watts--Strogatz topology. This is due to the enhanced connectivity of the Random Geometric Graph, which provides greater path diversity and mitigates network congestion during entanglement request processing.

\begin{table*}
\centering
\caption{Mean delay ($\mu$s) per request in different topology medium network}
\renewcommand{\arraystretch}{1.2}
\begin{tabular}{|p{2.3cm}|p{2.1cm}|p{1.8cm}|p{1.8cm}|p{1.5cm}|p{2.5cm}|p{1cm}|p{1cm}|}
\hline
Network Type & Dynamic Efficient & Dynamic FIFO & Static Efficient & Static FIFO & Success Enhancement & PPO & MILP \\
\hline
Watts-Strogatz & 179.38 & 180.15 & 221.45 & 384.45 & 336.41 & 205.79 & 177.22 \\
\hline
Random Geometric & 186.75 & 186.81 & 216.58 & 303.50 & 248.67 & 183.71 & 177.15 \\
\hline
\end{tabular}
\label{tab:topology_mean_delay}
\end{table*}

\begin{table*}
\centering
\caption{Number of successes in different topology medium network}
\renewcommand{\arraystretch}{1.2}
\begin{tabular}{|p{2.3cm}|p{2.1cm}|p{1.8cm}|p{1.8cm}|p{1.5cm}|p{2.5cm}|p{1cm}|p{1cm}|}
\hline
Network Type & Dynamic Efficient & Dynamic FIFO & Static Efficient & Static FIFO & Success Enhancement & PPO & MILP \\
\hline
Watts-Strogatz & 5204 & 5228 & 5301 & 5253 & 5408 & 5459 & 5429 \\
\hline
Random Geometric & 5341 & 5351 & 5369 & 5407 & 5527 & 5533 & 5519 \\
\hline
\end{tabular}
\label{tab:topology_num_success}
\end{table*}

\begin{table*}
\centering
\renewcommand{\arraystretch}{1.2}
\addtolength{\tabcolsep}{10pt}
\caption{Overall Average Capacity Utilization and Handling Rate}
\begin{tabular}{l|cc|cc}
\hline
\textbf{Method} & \multicolumn{2}{c|}{\textbf{Capacity Utilization}} & \multicolumn{2}{c}{\textbf{Handling Rate}} \\
 & Watts-Strogatz & Random Geometric & Watts-Strogatz & Random Geometric \\
\hline
Dynamic Efficient & 0.09 & 0.06 & 1.00 & 1.00 \\
Dynamic FIFO & 0.09 & 0.06 & 1.00 & 1.00 \\
Static Efficient & 0.09 & 0.06 & 0.95 & 0.96 \\
Static FIFO & 0.09 & 0.06 & 0.85 & 0.90 \\
Success Enhancement & 0.12 & 0.08 & 0.82 & 0.91 \\
PPO & 0.16 & 0.13 & 0.94 & 0.96 \\
MILP & 0.12 & 0.09 & 0.98 & 0.99 \\
\hline
\end{tabular}
\label{tab:capacity_handle}
\end{table*}

\section{Conclusion} 
\label{sec: conclusion}

This paper investigates resource allocation for entanglement distribution in multi-node, multi-user quantum networks with heterogeneous link qualities and diverse topologies. A multi-slot simulation framework incorporating request retry mechanisms is developed to evaluate resource allocation strategies that support concurrent entanglement distribution for multiple requests.

The results demonstrate clear trade-offs among request delay, entanglement success rate, capacity utilization, and request handling rate. Among the heuristic approaches, Dynamic Efficient achieves the lowest request delay but results in the fewest requests satisfying the fidelity threshold. Success Enhancement maximizes the entanglement success rate through adaptive multi-path allocation, at the cost of increased delay. Static Efficient provides a balanced trade-off, achieving lower delay than Success Enhancement while maintaining a higher success rate than Dynamic Efficient.

To jointly optimize request scheduling and multi-path resource allocation, a mixed-integer linear programming (MILP) approach is proposed to derive optimal resource assignments by maximizing a predefined objective function. Additionally, a PPO-based allocation method is introduced to learn adaptive resource allocation policies through reward-driven optimization. Both approaches achieve high request handling rates, thereby reducing request delays and improving the overall number of successful entanglement requests.

As for the computational complexity, there is a clear trade-off between scheduling performance and computational cost. The static methods have the lowest complexity and computation time, followed by the dynamic and Success Enhancement methods, and then the PPO for online inference. MILP provides the strongest overall allocation performance but has exponential worst-case complexity and the highest computation time, which may limit its use in large-scale or latency-sensitive deployments. The heuristic methods are therefore most suitable for rapid scheduling, while PPO offers a practical compromise between performance and online complexity.

Future work can be extended in several directions. First, future work will consider a more resource-abundant quantum network. Under this assumption, (1) multiple quantum channels per link can be exploited to simultaneously serve multiple requests instead of allocating repeated entanglement generation attempts to a single request; (2) fairness-aware multi-path allocation strategies can be investigated to improve resource sharing among concurrent requests; and (3) entanglement purification can be incorporated to enhance end-to-end entanglement fidelity, thereby improving the overall success probability of entanglement distribution. Second, dynamic network conditions, such as variations in resource availability, network topology, channel quality, and dynamic entanglement generation rate, could be taken into account. Under such dynamic environments, the computational complexity of resource allocation algorithms becomes increasingly important, as scheduling decisions must be made within limited time. Although this paper employs a pre-trained PPO model for inference, PPO is fundamentally an online policy optimization algorithm and therefore has the potential to continuously adapt to changing network conditions and topologies through ongoing learning. The results demonstrate the performance of PPO as one representative learning-based scheduler. Thirdly, a controlled comparison involving other learning approaches such as DQN, actor–critic variants and graph-based reinforcement-learning architectures remains an important direction for future work.

\section*{Acknowledgment}
This publication has emanated from research conducted with the financial support of Research Ireland under Grant number \text{13/RC/2077\_P2}. For the purpose of Open Access, the author has applied a CC BY public copyright licence to any Author Accepted Manuscript version arising from this submission.

\bibliographystyle{IEEEtran}
\bibliography{ref}

\EOD

\end{document}